\documentclass[acmtog]{acmart} 
\pdfoutput=1 
\usepackage{geometry}
\usepackage[utf8]{inputenc}
\usepackage{amsmath}
\usepackage{amsfonts}
\usepackage{soul}
\usepackage{microtype}
\usepackage{booktabs}
\usepackage{hyperref}
\usepackage{cleveref}
\usepackage{multirow}
\usepackage{graphicx}
\usepackage{dsfont}
\usepackage{pifont}
\usepackage{xcolor}
\usepackage{tikz}
\usepackage[skins]{tcolorbox}
\usepackage{wrapfig}
\usepackage{nicefrac}
\usepackage{bbm}
\usepackage{flushend}
\usepackage{tikz,pgfplots}
\usepackage{xspace}
\usepackage[ruled,vlined]{algorithm2e}
\usepackage{cancel}
\usepackage{adjustbox}
\usepackage{multirow}
\usepackage{tabularx}

\usepackage{overpic}  
\usepackage[pdftex,outline]{contour}

\DeclareGraphicsExtensions{.png,.jpg,.pdf,.ai,.psd}
\Crefname{figure}{{Fig.}}{{Figs.}}
\crefname{figure}{{Fig.}}{{Figs.}}

\definecolor{Red}{rgb}{1,0.45,0.45}
\definecolor{DarkRed}{rgb}{1,0,0}
\definecolor{Green}{rgb}{0.44,0.76,0.33}
\definecolor{Blue}{rgb}{0.46,0.7,1}
\definecolor{DarkGreen}{rgb}{0.0,0.6,0.0}
\definecolor{IntegralColor}{rgb}{1,0.645,0}
\definecolor{CVfunctionColor}{rgb}{1,0.07,0.57}
\definecolor{CVIntegralColor}{rgb}{1,0.4,0.7}
\definecolor{ExtSpaceColor}{rgb}{0.86,0,0.86}
\definecolor{TargetSpaceColor}{hsb}{0.45,0.87,0.81}
\definecolor{TargetMeasureColor}{rgb}{0.3,0.3,0.3}
\definecolor{RedClassColor}{hsb}{0.99,0.62,0.9}
\definecolor{BlueClassColor}{hsb}{0.61,0.6,0.9}
\definecolor{ProjectionColor}{rgb}{0.95,0.57,0.00}

\usepackage[outline]{contour}  
\usepackage{newfloat}  
\usepackage{xparse}
\newcommand{\undefinecolor}[1]{\expandafter\let\csname\string\color@#1\endcsname\undefined}
\DeclareDocumentCommand{\Outlined}{ O{black} O{white} O{0.55pt} m }{%
    \contourlength{#3}
    \contour{#2}{\textcolor{#1}{#4}}%
}

\hypersetup{draft} 

\usepackage[nolist]{acronym}
\begin{acronym}
\acro{MC}{Monte Carlo}
\acro{MIS}{Multiple importance sampling}
\end{acronym}

\newcommand{\ie}{i.\,e.,\xspace }

\newcommand{\Domain}{\Omega}

\newcommand\Coefficient{\mathrm{c}}
\newcommand\Basis{\phi}
\newcommand\numBasis{M}
\newcommand\Lnorm[1]{\mathcal{L}_{#1}}

\newcommand\BasisInnerProduct[2]{\mathrm{\Phi}({#1},{#2})}
\newcommand\BasisInnerProductMatrix{\mathrm{\Phi}}

\newcommand\Integral{F}

\newcommand\Integrand{f}
\newcommand\IntegrandPSS{\hat{f}}
\newcommand\PDF{p}

\newcommand\diff{\mathrm{d}}
\newcommand\CVfunction{{g}}
\newcommand\CVfunctionPSS{\hat{g}}
\newcommand\CVIntegral{G}
\newcommand\RIntegral{R}
\newcommand\Order[1]{O{#1}}

\newcommand\Estimator[1]{\langle #1 \rangle}

\newcommand\CVEstimator{\Estimator{\Integral}^*}

\newcommand\DEstimator{\Estimator{D}}
\newcommand\REstimator{\Estimator{R}}
\newcommand\FEstimator{\Estimator{\Integral}}

\newcommand\ExpectedOp[1]{\operatorname{E}\left[ #1 \right]}
\newcommand\VarianceOp[1]{\operatorname{Var}\left[ #1 \right]}
\newcommand\CovarianceOp[1]{\operatorname{Cov}\left[ #1 \right]}

\newcommand\Summation{\sum_{i=1}^{N}}

\usepgfplotslibrary{fillbetween}
\pgfplotsset{width=6cm, compat=1.10}
\pgfmathdeclarefunction{gaussian}{0}{\pgfmathparse{(1/2)*exp(-x^2/2)}}
\pgfmathdeclarefunction{poly2}{0}{\pgfmathparse{-1.65646886e-02*x^2+-1.19082167e-05*x+2.63495887e-01}}
\pgfmathdeclarefunction{poly4}{0}{\pgfmathparse{1.94354270e-03*x^4 -1.72247619e-06*x^3 -5.81931210e-02*x^2 +3.48720784e-05*x +3.67384898e-01}}

\AtBeginDocument{%
  \providecommand\BibTeX{{%
    \normalfont B\kern-0.5em{\scshape i\kern-0.25em b}\kern-0.8em\TeX}}}

\setcopyright{acmcopyright}
\copyrightyear{2022}
\acmYear{2022}

\acmJournal{TOG}
\acmVolume{41}
\acmNumber{4}
\acmMonth{7}

\setcopyright{acmlicensed}\acmJournal{TOG}
\acmYear{2022}\acmVolume{41}\acmNumber{4}\acmArticle{79}\acmMonth{7} \acmDOI{10.1145/3528223.3530095}

\begin{document}

\title{Regression-based Monte Carlo Integration}

\author{Corentin Sala\"{u}n}
\affiliation{%
  \institution{Max-Planck-Institut f{\"u}r Informatik}
  \city{Saarbr{\"u}cken}
  \country{Germany}
}
\email{csalaun@mpi-inf.mpg.de}

\author{Adrien Gruson}
\affiliation{%
  \institution{McGill University \& École de Technologie Supérieure}
  \city{Montreal}
  \country{Canada}
}
\email{adrien.gruson@gmail.com}

\author{Binh-Son Hua}
\affiliation{%
  \institution{VinAI Research}
  \country{Vietnam}
}
\email{binhson.hua@gmail.com}

\author{Toshiya Hachisuka}
\affiliation{%
  \institution{University of Waterloo}
  \city{Waterloo}
  \country{Canada}
}
\email{toshiya.hachisuka@uwaterloo.ca}

\author{Gurprit Singh}
\affiliation{%
  \institution{Max-Planck-Institut f{\"u}r Informatik}
  \city{Saarbr{\"u}cken}
  \country{Germany}
}
\email{gsingh@mpi-inf.mpg.de}

\renewcommand{\shortauthors}{C. Sala\"{u}n, A. Gruson, B-S. Hua, T. Hachisuka, G. Singh}

\begin{teaserfigure}
    \input{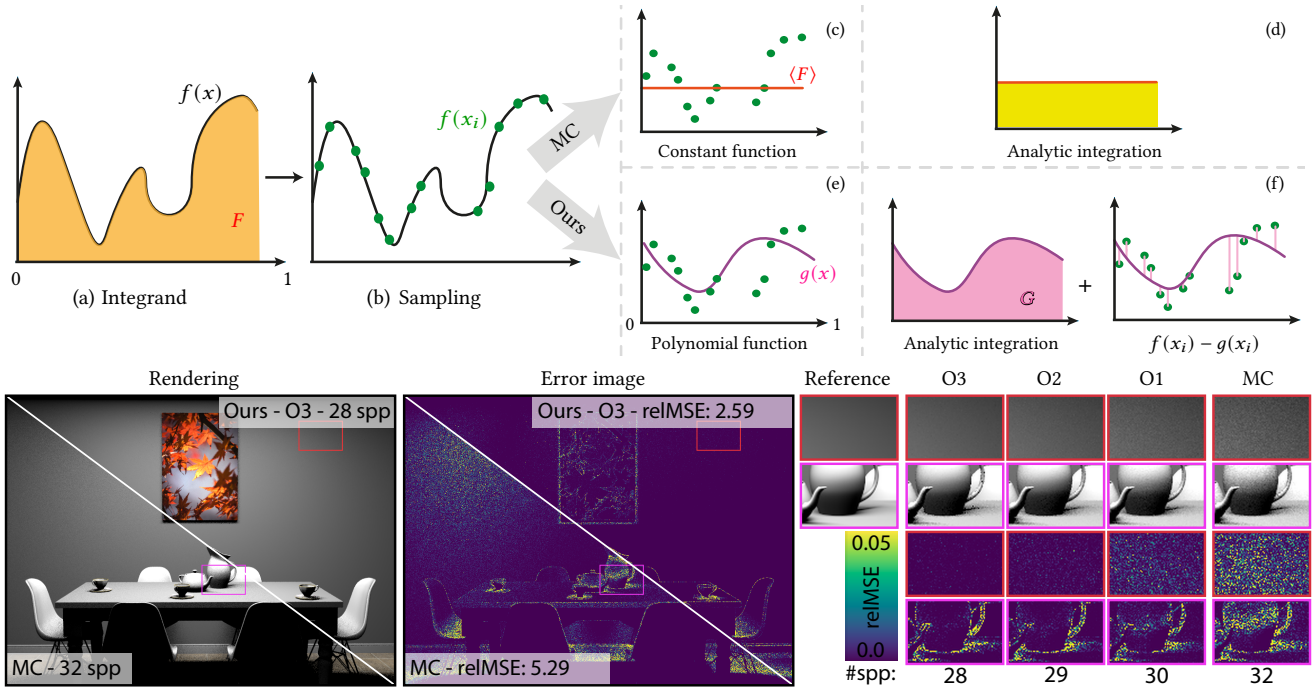}
  \caption{
        Given an integrand (a), we first sample (b) $f(x)$ as in Monte Carlo (MC) integration. 
        (c-d) Traditional MC estimator can be interpreted as fitting a  \emph{constant} model function to the sample values, with the integral of this constant function equals to $\Integral$.
        (e) We, instead, propose to use a \emph{non-constant} model function such as a polynomial, which 
        is then fitted to the sampled values. 
        (f) The resulting estimator is based on control variates; we add the analytical integral of the {model} function to MC integration of the difference between the original integrand and the {model} function. 
        The bottom row shows renderings and the corresponding error images to demonstrate the impact of our regression approach against the traditional MC integration. 
        The insets on the right compare our method with different orders ($O$x) of polynomials.  
        Our method has significant error reduction at equal time.
  }
  \label{fig:Teaser}
\end{teaserfigure}

\begin{abstract}

Monte Carlo integration is typically 
interpreted as an estimator of the expected value using stochastic samples. 
There exists an alternative interpretation in calculus where Monte Carlo integration can be seen as estimating a \emph{constant} function---from the stochastic evaluations of the integrand---that integrates to the original integral.
The integral mean value theorem states that this \emph{constant} function should be the mean (or expectation) of the integrand.
Since both interpretations result in the same estimator, little attention has been devoted to the calculus-oriented interpretation.
We show that the calculus-oriented interpretation actually implies the possibility of using a more \emph{complex} function than a \emph{constant} one to construct a more efficient estimator for Monte Carlo integration. 
We build a new estimator based on this interpretation and relate our estimator to control variates with least-squares regression on the stochastic samples of the integrand. 
Unlike prior work, our resulting estimator is \emph{provably} better than or equal to the conventional Monte Carlo estimator.
To demonstrate the strength of our approach, we introduce a practical estimator that can act as a simple drop-in replacement for conventional Monte Carlo integration. We experimentally validate our framework on various light transport integrals.
The code is available at \url{https://github.com/iribis/regressionmc}.

\end{abstract}



\maketitle

\section{Introduction}
\ac{MC} integration is a basic tool for many numerical integration problems.
In textbooks, \ac{MC} integration is typically interpreted as a statistical process where it samples the integrand and estimates the expected value by taking the average of these samples. 
While less common, there is another interpretation of \ac{MC} integration based on the mean value theorem in calculus.
The mean value theorem for definite integrals tells us that there exists a \emph{constant} that integrates exactly to the same solution as the given definite integral (see~\cref{fig:Teaser}). 
This {constant}, by definition, is equal to the mean of the integrand, which can be again estimated by taking the average of random samples, and thus both interpretations result in the same estimator.
This alternative interpretation, however, raises interesting questions: \emph{Can we use a non-constant function to construct a MC estimator? If so, what happens?}
These questions do not appear in the statistics-oriented interpretation as the expected value is fundamentally tied to the idea of samples' average.

We propose a new \ac{MC} estimator that takes the same input as \ac{MC} integration (see~\cref{fig:Teaser}b) with no extra assumptions. 
Instead of merely taking the average to find a {constant} function (\cref{fig:Teaser}c), we propose to fit a \emph{non-constant} model function (\cref{fig:Teaser}e) to the samples via least-squares regression and utilize its analytical integral to define an estimator (\cref{fig:Teaser}f). 
We show that our approach can be formulated as the method of control variates, where control variates are obtained via least-squares regression.

While the use of regression in control variates is not new, we show, for the first time how regression, control variates, and \ac{MC} integration are closely coupled. 
 \ac{MC} integration is a special case under our formulation where a constant function is the solution of regression.
Our approach is provably shown to be better than or equivalent to \ac{MC} integration; in the worst case, it will revert back to conventional \ac{MC} integration as long as a constant function is included in regression. 
This theoretical guarantee has not been explored in prior work on regression-based control variates.
To summarize our contributions:
\begin{itemize}
    \item We introduce a novel formulation of \ac{MC} integration which connects least-squares regression, control variates, and conventional \ac{MC} integration.
    \item We prove that our regression-based formulation leads to a class of estimators that are better than or equal to conventional \ac{MC} integration in terms of variance.
    \item We develop a practical algorithm based on our formulation and polynomials, which can work as drop-in improvement over conventional \ac{MC} integration.
    \item We demonstrate the effectiveness of our algorithms over light transport integral estimations by performing experiments over multiple dimensions (2D to 15D).

\end{itemize}
%

\section{Related work}
%
\subsection{Numerical integration} 
Integration is a classical subject in numerical analysis~\cite{trefethen2012approximation,davis2013methods}. 
The calculus-oriented approach has lead to deterministic quadrature techniques such as a Riemann sum.
The quadrature techniques sample the integrand at predetermined points and approximate the given integrand using an analytically integrable function (e.g., quadrangles, trapezoids, and certain basis functions).
The calculus-oriented approach, in general, suffers from the exponential growth of the number of samples for higher dimensional functions.
A statistics-oriented approach, \emph{\ac{MC} integration}, performs a similar summation but instead evaluates the integrand at stochastically sampled locations. 
This stochastic nature of sample generation extends well to higher dimensions~\cite{Owen:2013:Monte,Niederreiter:1992:Random} without the exponential growth of the number of samples, which has led to its wide adoption in the rendering community. 
We take a step back from this trend and show how a calculus-oriented reformulation of \ac{MC} integration can lead to more efficient numerical integration methods.

\subsection{Control variates} 
Control variates (CVs)~\cite{glynn:2002:controlVariates,Loh:1995:ControlVariates} are variance reduction techniques which utilize \ac{MC} integration of the difference between a given integrand $\Integrand$ and another function $\CVfunction$ to perform numerical integration. 
This difference estimate is added to the analytical integral of the function $\CVfunction$ to estimate the integral of $\Integrand$. 
The improvement over conventional \ac{MC} integration is dependent on how well $\CVfunction$ approximates $\Integrand$, and the improvement is \emph{not guaranteed} in general.

\citet{Owen:2000:Safe} discussed a practical approach to perform importance sampling with control variates for variance reduction. They showed that adaptive control variates using a mixture density cannot be \emph{much worse} (\ie can only be slightly worse) than conventional \ac{MC} using the best mixture component. Our approach, on the other hand, is \emph{provably never worse} than conventional \ac{MC} using the same PDF (shown in~\Cref{subsec:reduce_mc} and \Cref{eq:proof_mce_ours}). \citet{Fan:2006:Optimizing} used the method of Owen and Zhou in rendering with an additional approximation.

Multi-level Monte Carlo~\cite{Heinrich:2001:MultilevelMC,Keller:2001:Hierarchical} shows how low resolution renderings can be used as CVs for high-resolution synthesis. 
\citet{Rousselle:2016:Imagespace} shows a connection between CVs and the reconstruction step in gradient-domain rendering~\cite{hua2019survey}. 
Recently, \citet{Kondapaneni:2019:Optimal} showed a connection between weights in multiple importance sampling~\cite{Veach:1995:Optimally} and CVs. 
We show connections among control variates, least-squares regression, and \ac{MC} integration to have a provable variance reduction. 
We will discuss more related work~\cite{nakatsukasa2018approximate,pajot2014globally,Owen:2000:Safe,crespo2020primaryspace,hickernell2005control,rubinstein1985efficiency} later once we introduced our approach.

\paragraph{Construction of control variates in rendering}
Constructing control variates for complex integrands in light transport simulation is a challenging task. 
In rendering, the simplest form of control variate uses the constant ambient term to approximate incoming illumination~\cite{Lafortune:1994:Ambient}. 
This, however, does not work well in complex lighting conditions. 
\citet{Clarberg:2008:Exploiting} used visibility correlations to design control variates for direct illumination. \citet{vevoda2018bayesian} used bayesian regression to construct a light clustering. For further improvement, a control variate function is derived from the respective estimated contribution of each cluster at any shading point.
\citet{kutz2017spectral} proposed a method of control variate extension for heterogenous participating media rendering. It is based on the decomposition of the medium into a control and a residual components. These two parts can be sampled separately.
The control variate has also been used by \citet{belcour2018integrating} to extend their method of integration of spherical harmonic expansions over polygonal domains to non-analytic integrands.
Neural control variates~\cite{Mueller:2020:NeuralCV} fit a neural network to define control variates. 
They also proposed to simultaneously train another network to importance sample the difference function. 
The resulting estimator shows remarkable improvements, but with a significant computational overhead and changes to the sampling process that have to be accelerated by utilizing a high-performance multi-GPU workstation. 
Recently, \citet{subr2021q} propose a neural network Q-NET that uses a neural proxy as a control variate.
Our control variates are constructed from the samples within each pixel using least squares regression. 

The expected improvement from these control variates is limited by construction due to the lack of complexity captured by these control variates.
It is thus reasonable to assume that control variates usually do not approximate the integrand well in light transport simulation.
Recognizing this limitation, our approach is designed to be provably better than MC integration as long as least-squares regression can find \emph{any} non-constant solution.
A major deviation from prior work is that we show how regression is closely connected to control variates and \ac{MC} integration. Many prior work on control variates instead focus on introducing a sophisticated model function for control variates.
This conceptual difference allows us to realize that we can use a \emph{significantly simpler} model function, such as polynomials, and still provably outperforms \ac{MC} integration.
We demonstrate that our estimators are already practical without any special optimization for regression (e.g., use of high performance hardware like GPUs) and without any change to the sampling process.

\subsection{Regression in rendering}
Regression has been applied for adaptive sampling and reconstruction in rendering~\cite{Zwicker:2015:Recent}. 
\citet{Moon:2014:Adaptive} proposed to locally approximate the unknown image function by locally regressing a low-order polynomial on auxiliary buffers (normals, depth, textures). 
To alleviate the computational overhead of the method, \citet{Moon:2015:Adaptive} used a iterative linear model to simultaneously reconstruct multiple pixels. These methods control the filtering bandwidths locally to increase numerical accuracy. 
\citet{Bitterli:2016:Nonlinearly} analyze various zero-order and first-order regression models and use auxiliary data just to fit the data and considers only the pixel color to compute the regression weights.
Our framework is largely agnostic to how regression is implemented, and we demonstrate multiple implementations as examples.
The regression part of our framework can thus utilize findings in those previous.

\section{Two interpretations of \ac{MC} integration}
Let us motivate our main question by recapitulating the two different interpretations of MC integration.
Suppose that we want to find a definite integral $\Integral$ of a non-negative function $\Integrand(x)$ over the domain $\Omega$ where 
\begin{align}
    \Integral = \int_{\Omega} \Integrand(x) \diff x.
\end{align}
In many practical problems, it is difficult to find the analytical solution to the above.
In light transport simulation, the function $\Integrand(x)$ is defined as the measurement contribution function for a light transport path $x$~\cite{Veach:1997:Robust}. 
The definite integral in this case is the solution to light transport simulation, which is generally unavailable in an analytical form.

Rather than trying to find an analytical solution, \ac{MC} integration uses a numerical estimator $\FEstimator{}$ where
\begin{align}
    \label{eq:mc_estimator}
    \Integral \approx \FEstimator = \frac{1}{N} \sum_{i=1}^{N} \frac{\Integrand(x_i)}{\PDF(x_i)}. 
\end{align}
The variables $x_i$ are $N$ independent random samples distributed according to the probability density function $\PDF(x)$ which satisfies $\PDF(x) \neq 0$ whenever $\Integrand(x) \neq 0$ in the domain $\Domain$. 
This estimator $\FEstimator{}$ is said to be \emph{unbiased} in the sense that its expected value is exactly equal to the integral $\Integral$:
\begin{align}
    \ExpectedOp{\FEstimator} = \frac{1}{N} \sum_{i=1}^{N} \ExpectedOp{\frac{\Integrand(x_i)}{\PDF(x_i)}} =  \frac{1}{N} \sum_{i=1}^{N} \int_{\Omega}  \frac{\Integrand(x)}{{p(x)}} {p(x)} \diff x = \Integral.
\end{align}
\paragraph{Change of variables}
The same estimator can also be redefined in a unit hypercube $U$ of uniformly distributed random numbers $u_i$ that are used for generating samples $x_i$.
Suppose that we have a mapping $x = \Phi(u)$ where the Jacobian $|\diff x/ \diff u| = 1 / \PDF(x)$.
By change of variables, we can rewrite the integral into 
\begin{align}
    \Integral =  \int_{\Omega}  \Integrand(x) \diff x = \int_{U}  \Integrand(\Phi(u)) \left| \frac{\diff x}{\diff u} \right| \diff u = \int_{U} \IntegrandPSS(u) \diff u,
\end{align}
where $\IntegrandPSS(u) = \frac{\Integrand(\Phi(u))}{\PDF(\Phi(u))}$, and rewrite the estimator $\FEstimator$ into
\begin{align}
    \Integral \approx \FEstimator= \frac{1}{N} \sum_{i=1}^{N} \frac{\IntegrandPSS(u_i)}{\PDF(u_i)} = \frac{1}{N} \sum_{i=1}^{N} \IntegrandPSS(u_i).
\end{align}
In other words, by properly redefining the integrand as above, one can always consider having uniformly distributed samples $u_i \sim \PDF(u) = 1$ in this unit hypercube $U$, instead of samples $x_i \sim \PDF(x)$ in the domain $\Omega$ which are generally not uniformly distributed.
This definition is known as as primary sample space~\cite{Kelemen:2002:Simple} in rendering.
We use this formulation in the rest of the paper. One can generally convert Equation 1 to Equation 4 as long as change of variables from $x$ to $u$ is possible.

\subsection{Statistic-oriented interpretation}
The process of \ac{MC} integration can be seen as an estimation of the expected value of $\IntegrandPSS(u)$ since we have  
\begin{align}
    \Integral = \int_{U} \IntegrandPSS(u) p(u) \diff u
    = \ExpectedOp{\IntegrandPSS(u)} \approx \frac{1}{N} \sum_{i=1}^{N} \IntegrandPSS(u_i) = \FEstimator 
\end{align}
where $u$ is distributed according to $\PDF(u)=1$.
Under this interpretation, one can also derive an expression of the expected squared errors of $\FEstimator$ as
\begin{align}
    \ExpectedOp{\left( \FEstimator{} - \Integral \right)^2} = \VarianceOp{\FEstimator{}} =  \frac{1}{N} \VarianceOp{\IntegrandPSS(u)}
\end{align}
where we used the unbiasedness of the estimator $\ExpectedOp{\FEstimator{}} = \Integral$.
It is a well known expression that shows that the expected squared error of \ac{MC} integration is proportional to the variance of $\IntegrandPSS(u)$ and inversely proportional to the number of samples $N$.

\subsection{Calculus-oriented interpretation}
One can also interpret \ac{MC} integration as a two-step process; \emph{replacement} of the integrand $\IntegrandPSS$ by a model function $\CVfunctionPSS$ and succeeding \emph{analytical} integration of $\CVfunctionPSS$.
Let us consider a constant function $\CVfunctionPSS(u) = c$ as the model function.
To keep its integral $G$ equal to the original integral $F$, we would like to find a value for $c$ such that  
\begin{align}
    \Integral = \int_{U} \IntegrandPSS(u) \diff u = \int_{U} \CVfunctionPSS(u) \diff u = G.
\end{align}
The integral mean value theorem in calculus shows that there exist such a constant $c$ for a given definite integral.
Since the function $\CVfunctionPSS(u)$ can be analytically integrated, we have 
\begin{align}
   F = G = \int_{U} \CVfunctionPSS(u) \diff u = c |U|  \quad \therefore c  = \frac{1}{|U|} \int_{U} \CVfunctionPSS(u) \diff u
\end{align}
where $|U| = 1$ is the volume of the unit hypercube $U$. 
Therefore, the constant $c$ can be derived as the \emph{average} of $\IntegrandPSS(u)$ since 
\begin{align}
   c  = \frac{1}{|U|} \int_{U} \CVfunctionPSS(u) \diff u = \frac{1}{|U|} \int_{U} \IntegrandPSS(u) \diff u
\end{align}
where the last expression is the definition of the average of $\IntegrandPSS(u)$ over the unit hypercube $U$. 
Note that there is no approximation introduced until this point.
Since $c$ is defined as the average of $\IntegrandPSS(u)$, we can estimate $c$ by taking the average of samples;
\begin{align}
   c  \approx \frac{1}{N} \sum_{i=1}^{N} \IntegrandPSS(u_i) = \FEstimator.
\end{align}
This interpretation thus leads to the same estimator as the statistic-oriented interpretation since 
\begin{align}
    \Integral = \int_{U} \CVfunctionPSS(u) \diff u = \int_{U} c \diff u \approx \int_{U} \left( \frac{\FEstimator}{|U|}  \right) \diff u = \FEstimator. 
\end{align}
Under this interpretation, \ac{MC} integration can be seen as the process of first estimating a function $\CVfunctionPSS(u) = c$, and then analytically integrating (the approximation of) the function $\CVfunctionPSS(u)$. 

\subsection{Problem statement}
One might see that the calculus-oriented interpretation is  rather redundant at first, since the end result is the same. However, this interpretation actually points out the existence of $\CVfunctionPSS$ in the definition of the estimator, and also the fact that \ac{MC} integration is derived by using a constant function $\CVfunctionPSS = c$.
This existence of $\CVfunctionPSS$ is not apparent in the statistics-oriented interpretation.

Our main question is whether making $\CVfunctionPSS$ \emph{non-constant} is possible 
and whether it is beneficial to consider such $\CVfunctionPSS$ to begin with.
We show that such a $\CVfunctionPSS$ can be defined based on \emph{control variates} and \emph{least-squares regression} of $\CVfunctionPSS$ to $\IntegrandPSS$. 
Our formulation shows that conventional \ac{MC} integration is a special case where regression \emph{happens to return} a constant function for $\CVfunctionPSS$. 
We also show that considering a more complex $\CVfunctionPSS$ is \emph{provably better} than or equal to simply having a constant $\CVfunctionPSS$ in terms of its expected error.

\section{Regression-based \ac{MC} integration}
\label{sec:regression_estinator}
We introduce a novel formulation of \ac{MC} integration, \emph{regression-based \ac{MC} integration}, which allows us to consider any analytically integrable functions for $\CVfunctionPSS$, in place of the constant function.
Fig.~\ref{fig:Teaser} illustrates our approach for a 1D function.

We start by considering an arbitrary model function $\CVfunctionPSS(u)$ which can be analytically integrated. 
Unlike the calculus-oriented interpretation, our function $\CVfunctionPSS(u)$ can be non-constant and the integral $\CVIntegral$ \emph{does not need to be} equal to the original integral $\Integral$. 
We can use such $\CVfunctionPSS(u)$ by adding the \emph{difference} from the original integral as
\begin{align}
    \Integral = \CVIntegral + (\Integral - \CVIntegral) 
\end{align}
where $(\Integral - \CVIntegral)$ is estimated by \ac{MC} integration;
\begin{align}
    \Integral - \CVIntegral 
    \approx \frac{1}{N} \sum_{i=1}^{N} \left( \IntegrandPSS(u_i) - \CVfunctionPSS(u_i) \right) = \DEstimator.
\end{align}
The final estimator is defined as a sum of two terms where
\begin{align}
    \label{eq:control_variate_estimator}
    \Integral \approx \CVEstimator = \CVIntegral + \DEstimator.
\end{align}
The first term $\CVIntegral$ is analytically evaluated for this given $\CVfunctionPSS(u)$ in a deterministic manner.
This estimator is not new and is known as the method of \emph{control variates}.
The function $\CVfunctionPSS(u)$ is called the control variate which is typically \emph{given} by a user. 
In our method, we find $\CVfunctionPSS(u)$ by (least-squares) \emph{regression}. 

Control variates are \emph{unbiased} regardless of  $\CVfunctionPSS(u)$. 
Since $\CVIntegral$ is deterministic, the expected squared error of $\CVEstimator$ is given by the difference estimator
\begin{align}
     \label{eq:control_variate_error}
     \ExpectedOp{\left( \CVEstimator - \Integral \right)^2}  =  \VarianceOp{\DEstimator} = \frac{1}{N} \VarianceOp{\IntegrandPSS(u) - \CVfunctionPSS(u)}. 
\end{align}
Control variates thus lead to a smaller expected square error compared to conventional \ac{MC} integration \emph{only} when $\operatorname{Var}[\IntegrandPSS(u) - \CVfunctionPSS(u)] < \operatorname{Var}[\IntegrandPSS(u)]$. 
Error reduction is thus \emph{not guaranteed} and it is widely recognized so~\cite{hickernell2005control}.

Our key contribution is to show that least-squares regression not only helps us automatically finding $\CVfunctionPSS(u)$ but also \emph{provably} reduces the expected squared error  \emph{when a constant function is included in regression}. 

As we discuss later, a constant function has been usually ignored in (adaptive) control variates. Our work is the first to show that this combination of the least-square regression and inclusion of a constant function results in the provable improvement with a proper reduction to MC integration.

\begin{figure*}[t!]
    \centering
    \includegraphics[width=\textwidth]{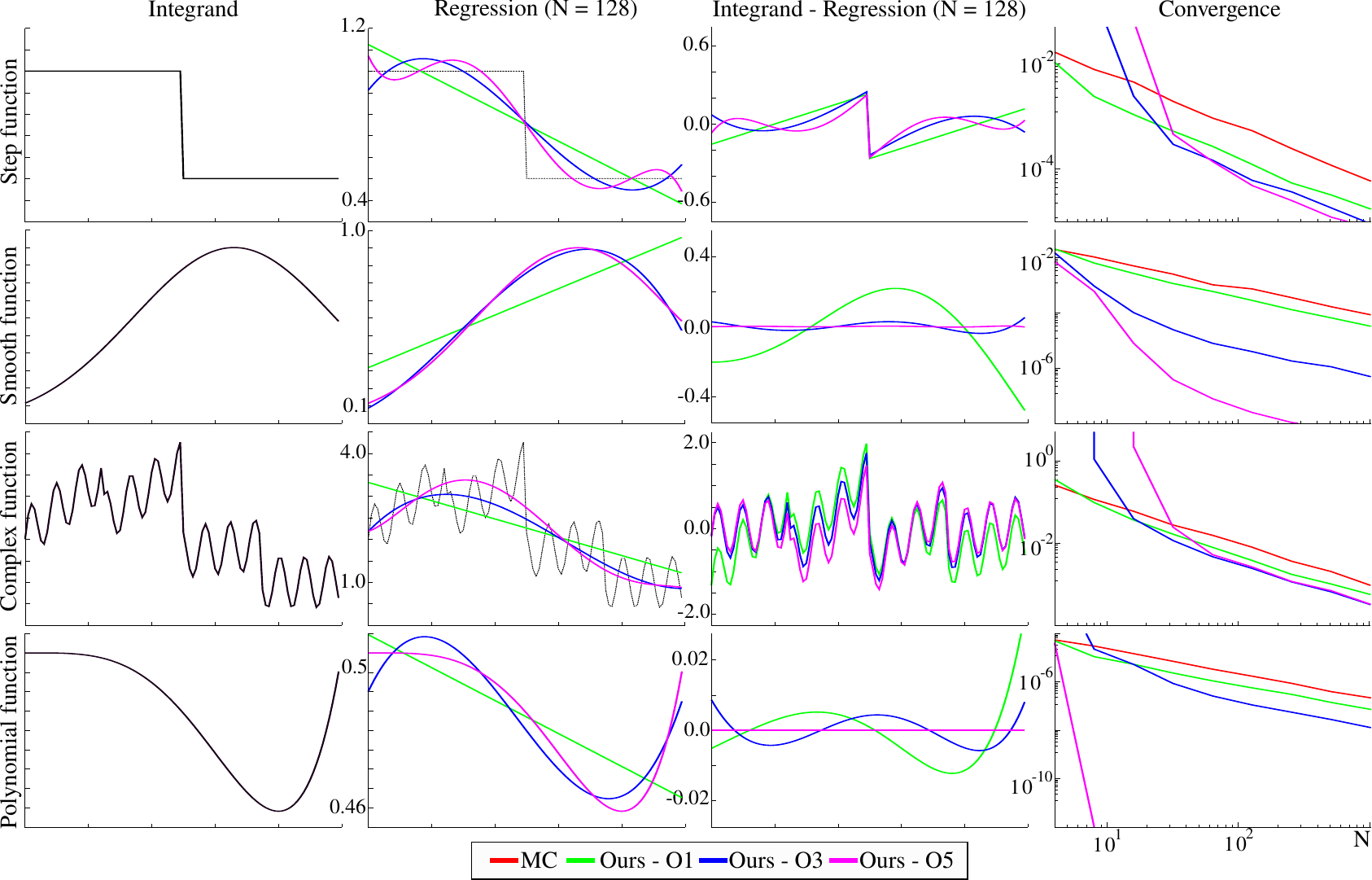}

    \caption{ Regression-based Monte Carlo integration in 1D: a step function, a shifted Gaussian, a high-frequency function, and a polynomial.
    Our method is no worse than \ac{MC} integration. It is not required that the model function well approximates the integrand for our method to work (e.g., the step function and high-frequency function case where our model function is very different from the integrand). When the model function is a good fit (e.g., the shifted Gaussian, polynomial function case), the error can be reduced dramatically.
    }
    \label{fig:1d_experiment}
\end{figure*}

Fig.~\ref{fig:1d_experiment} showcases our method for simple 1D integration problems. 
The labels $O1$, $O3$, and $O5$ in our method mean different orders of polynomials used for $\hat{g}$ as we explain later. 
The convergence plots show that, after a certain number of samples, our method is no worse than \ac{MC} integration regardless of the order of polynomials. 
It is also worth noting that regression might \emph{not} well approximate the integrand (e.g., step and complex cases). 
The method of control variates alone with such a bad approximation could \emph{increase} the error.
Even in such cases, our method still outperforms \ac{MC} integration once regression converges after some numbers of samples. 
When the integrand can be modeled exactly by $\hat{g}$, as shown in the last row where the integrand is a 5th-order polynomial, our method dramatically reduces numerical error compared to \ac{MC} integration.
We now explain how our estimator ${\langle F \rangle}^*$ achieves this reduction of errors by showing its connections to control variates and \ac{MC} integration. 

\subsection{Least-squares regression of $\CVfunctionPSS(u)$}
We model $\CVfunctionPSS(u, \theta)$ as a parametric function with the parameters $\theta = (\Coefficient_0, \cdots, \Coefficient_M)$. 
We define the \emph{residual} for $\theta$ as
\begin{align}
    \label{eq:residual_cont}
    \RIntegral(\theta) = \int_U \left( \IntegrandPSS(u) - \CVfunctionPSS(u, \theta) \right)^2 \diff u.
\end{align}
Let us then consider regression of $\CVfunctionPSS(u)$ to $\IntegrandPSS(u)$ such that the residual $\RIntegral(\theta)$ is minimized.
Since analytical solution to the integral in $ \RIntegral(\theta)$ is usually not available, we first consider using $N$ random samples $u_i$ to approximate $\RIntegral(\theta)$ 
\begin{align}
    \label{eq:residual}
    \RIntegral(\theta) \approx \frac{1}{N} \sum_{i=1}^{N} \left( \IntegrandPSS(u_i) - \CVfunctionPSS(u_i, \theta) \right)^2 = \REstimator(\theta), 
\end{align}
which is equivalent to \ac{MC} integration of the function $(\IntegrandPSS(u) - \CVfunctionPSS(u, \theta))^2$ using the estimator $\REstimator(\theta)$.
This unbiased approximation thus converges to $\RIntegral(\theta)$ as we increase the number of samples.
This expression of $\REstimator(\theta)$ can also be seen as the squared $\Lnorm{2}$ norm of the differences between $\IntegrandPSS(u_i)$ and $\CVfunctionPSS(u_i, \theta)$ on the samples $u_i$.
The solution to this regression can be obtained via \emph{least-squares regression} of $\CVfunctionPSS(u, \theta)$ to $N$ pairs of $(u_i, \IntegrandPSS(u_i))$.

\subsection{Connection to control variates}
The original definition of control variates considers a given function $h(u)$ and a scalar parameter $\alpha$ such that
\begin{align}
    \label{eq:control_variate_form}
    \Integral = \int_{U} \IntegrandPSS(u)  \diff u = \alpha H + \int_U ( \IntegrandPSS(u) - \alpha h(u)) \diff u
\end{align}
where $H = \int_U h(u) \diff u$.
Given this definition, it is known that  
\begin{align}
    \alpha = \CovarianceOp{\IntegrandPSS,h}/\VarianceOp{h}
\end{align}
minimizes the variance of $\CVEstimator$.
This approach can be seen as a special case of our formulation where $\theta = \alpha$ and $\CVfunctionPSS(u, \theta) = \alpha h(u)$.

Let us analyze the error of $\CVEstimator$ when $\CVfunctionPSS(u, \theta)$ is given by least-squares regression.
Since the expected squared error of $\CVEstimator$ is equal to the variance of $\IntegrandPSS - \CVfunctionPSS$ (\Cref{eq:control_variate_error}), we have
\begin{align}
    &\VarianceOp{\IntegrandPSS(u) - \CVfunctionPSS(u, \theta)}  =\ExpectedOp{\left( \IntegrandPSS(u) - \CVfunctionPSS(u, \theta) \right)^2} - \ExpectedOp{ \IntegrandPSS(u) - \CVfunctionPSS(u, \theta)}^2 \nonumber\\
    &= \int_U \left( \IntegrandPSS(u) - \CVfunctionPSS(u, \theta) \right)^2 \diff u - (F - G)^2 = \RIntegral(\theta) - (F - G)^2.
\end{align}
where we used the fact that $\PDF(u) = 1$ in the primary sample space and substituted $\RIntegral(\theta)$.
The expected squared error of the estimator is thus bounded as
\begin{align}
    \label{eq:upper_bound_residual}
    \ExpectedOp{\left( \CVEstimator - \Integral \right)^2} = \frac{1}{N}\left(\RIntegral(\theta) - (F - G)^2\right) \leq \frac{1}{N} \RIntegral(\theta).
\end{align}
We can see that, by defining the control variate $\CVfunctionPSS(u)$ via least-squares regression, one can also minimize (the bound of) the expected squared error of the estimator. 
In practice, we cannot directly minimize $\RIntegral(\theta)$, but we minimize its estimate $\REstimator(\theta)$, so this relationship does not exactly hold for regression with a finite number of samples as one can also see in Fig.~\ref{fig:1d_experiment}.
Our experiments, however, show that minimizing $\REstimator(\theta)$ works well in practice when $N$ is not too small.

Note also that control variates directly minimize the variance by setting $\alpha$, while we minimize only its bound by least-squares regression.

While it is tempting to consider more general regression to directly minimize the variance in our method, it is the use of least-squares regression which allowed us to reveal clear connections among least-squares regression, control variates, and \ac{MC} integration as we see in the followings.

\subsection{Reduction to Monte Carlo integration}
\label{subsec:reduce_mc}
It is now trivial to show that our approach reduces to \ac{MC} integration when regression happened to return a constant function $\CVfunctionPSS(u) = c$ as its best fit. 
Substituting $\CVfunctionPSS(u) = c$ to \Cref{eq:control_variate_estimator},
\begin{align}
    \CVEstimator = c|U| + \frac{1}{N} \sum_{i=1}^{N} \left( \IntegrandPSS(u_i) - c \right) =  \frac{1}{N} \sum_{i=1}^{N} \IntegrandPSS(u_i) = \FEstimator.
\end{align}
It shows that, when $\CVfunctionPSS(u) = c$ is the solution to regression, our approach reduces to \ac{MC} integration.

In this case, since we have $F = G$, the upper-bound in \Cref{eq:upper_bound_residual} becomes exact and 
\begin{align}
    \ExpectedOp{\left( \CVEstimator - \Integral \right)^2} = \ExpectedOp{\left( \FEstimator - \Integral \right)^2}  =  \RIntegral(c) 
\end{align}
where $\RIntegral(c)$ is the residual for $\CVfunctionPSS(u) = c$.

Based on this reduction to Monte Carlo integration, we can show for the first time that the expected squared error of this approach is \emph{provably} better than or equal to \ac{MC} integration. 
Deceivingly simple, but the important difference from prior work is that we include a constant function in least-squares regression by adding $c$ to $\CVfunctionPSS(u)$ and we minimize the residual $\RIntegral(\theta)$.
This allows us to have a "safe-guard" against performing worse than \ac{MC} integration, when combined with least-squares regression.
Since $\RIntegral(\theta)$ is minimized for a model function including a constant,  
\begin{align}
    \label{eq:proof_mce_ours}
    \RIntegral(\theta) \leq \RIntegral(c).
\end{align}
It thus follows that
\begin{gather}
    \ExpectedOp{\left( \CVEstimator - \Integral \right)^2} \leq \frac{1}{N} \RIntegral(\theta) \leq \frac{1}{N} \RIntegral(c) = \ExpectedOp{\left( \FEstimator - \Integral \right)^2} \nonumber\\
    \therefore \ExpectedOp{\left( \CVEstimator - \Integral \right)^2} \leq \ExpectedOp{\left( \FEstimator - \Integral \right)^2}.
\end{gather}
The two errors become equal only when $\RIntegral(\theta) = \RIntegral(c)$, that is, our estimator exactly reduces to \ac{MC} integration when a constant function is the solution to regression. 
This inequality also shows that, when a constant function is not the solution, it will always lead to a better or equivalent estimator $\CVEstimator$ than \ac{MC} integration $\FEstimator$.
Our estimator is thus \emph{provably better than or equal to} \ac{MC} integration in terms of its expected squared errors.

In practice, since the solution to least-squares regression is not exact for any finite $N$, this theoretical property of reduction of errors is not guaranteed for any finite $N$. 
Nonetheless, our numerical results demonstrate that reduction of error happens and stabilize well even for a practical range of $N$.

\paragraph{Discussion}

While regression is often used as an approach to find a control variate, none of the prior work shows how including a constant function in regression allows us to unify control variates and \ac{MC} integration. For example, a very thorough discussion on control variates \cite[Section~4]{hickernell2005control} immediately dismisses a constant control variate as redundant since it will not reduce variance at all.

Recognizing a problem of control variates that it could perform worse than \ac{MC} integration, \citet{pajot2014globally} proposed switching to conventional \ac{MC} when the estimated variance of adaptive control variate is determined to be larger. Our approach does not need this explicit switching, as it is automatically done as a result of least-squares with theoretical guarantee. 
Although our estimator looks similar to a general mixture estimator by~\citet{Owen:2000:Safe}, their estimator does not revert back to \ac{MC} integration. 
The difference becomes apparent when we fit a mixture density including a constant function. 
When regression returns a constant as the solution, the method of~\citeauthor{Owen:2000:Safe} also changes the sampling density to a constant.
On the other hand, our method reduces to \ac{MC} integration while keeping the sampling density as the given mixture density. The two estimators are thus fundamentally different.

Some existing methods~\cite{rubinstein1985efficiency,Rousselle:2016:Imagespace} use multiple different control variate functions to take advantage of their respective strengths. \citet{rubinstein1985efficiency}, in particular, shows that the variance reduction depends on the \emph{optimal} number of control variates used in the estimator. 
Our work is orthogonal to this work.

\section{Polynomial-based estimators}
\label{sec:implementation}
While our formulation supports arbitrary integrable functions for $\CVfunctionPSS$, we elaborate a practical implementation with a polynomial function for $\CVfunctionPSS$ to demonstrate the strength of our method. In general, a polynomial function can be written as 
\begin{align}
    \label{eq:function_approximation}
    \CVfunctionPSS(u) = \sum_{q=0}^{\numBasis} \Coefficient_q \Basis_q(u)\,
\end{align}
with monomials $\Basis$ defining a polynomial. 
We can analytically integrate $\CVfunctionPSS$ as the weighted sum of the integrals of monomials as
\begin{align}
    \label{eq:approximate_integrate}
    \int_{U} \CVfunctionPSS(u) \diff u
    =   \sum_{q=0}^{\numBasis} \Coefficient_q \int_{U}  \Basis_q(u) \diff u.
\end{align}

The advantage of polynomials is that it is simple and easily generalizable for different dimensions with different orders of polynomials. 
Note that we do not claim that polynomials are optimal as control variates. 
The integrand $\IntegrandPSS$ is, in general, far more complex than a polynomial, and it will not be well approximated by a polynomial in many cases.
Using polynomials in the conventional control variates might be seen as a bad approach because of this reason. 

What is important here is that a polynomial is still \emph{more complex} than a constant and includes a constant. 
Our formulation thus provably guarantees improvement over \ac{MC} integration in this case. 
Therefore, even with a lower-order polynomial, it is expected that our method either outperforms or at least performs equally to MC integration.
We explain two different approaches for regression.

\begin{algorithm}[t]
\SetAlgoLined
 $f \leftarrow \{\}$; $u \leftarrow \{\}$; $\BasisInnerProductMatrix_0 \leftarrow 0^{M \times M}$; $b_0 \leftarrow 0^{M \times 1}$\; 
 \For{$i \leftarrow 1$ \KwTo $N$}{
  $u_i \leftarrow$ random()\;
  $f_i \leftarrow \IntegrandPSS(u_i)$\;
  $(\BasisInnerProductMatrix_{i+1}, b_{i+1}) \leftarrow$ update\_system($\BasisInnerProductMatrix_i$, $b_i$, $u_i$, $f_i$)\; 
 }
 $\theta_{\mbox{min}} \leftarrow$ solve\_linear\_system($\BasisInnerProductMatrix_{N+1}$, $b_{N+1}$)\;
 \Return $G(\theta_{\mbox{min}}) + \frac{1}{N}(\sum_{i=1}^{N} f_i -  \CVfunctionPSS(u_i, \theta_{\mbox{min}}))$\;
\caption{\label{algo:DirectMatrixSolver}Direct matrix estimator.} 
\end{algorithm}
\begin{figure*}[t!]
    \centering
    \includegraphics[width=0.99\textwidth]{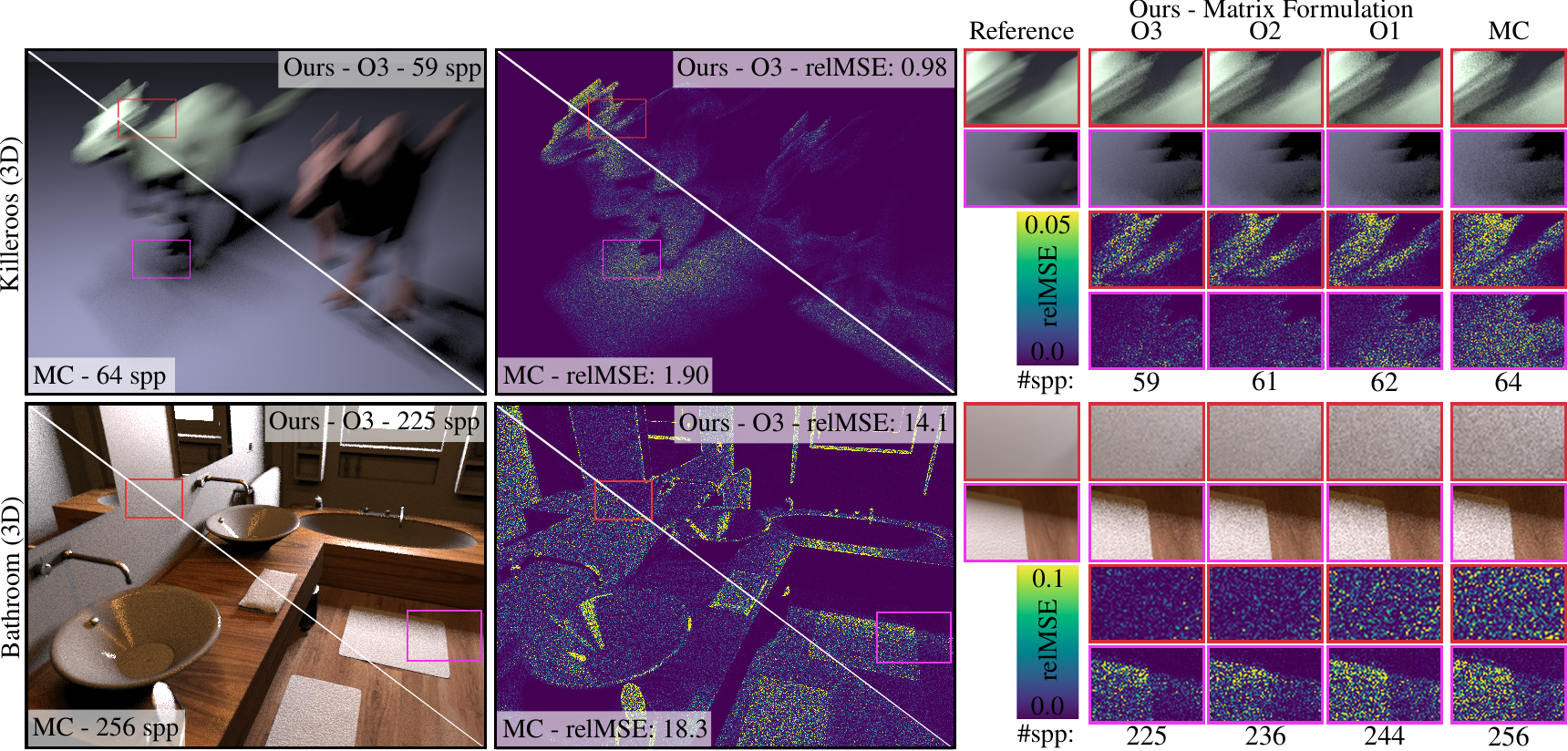}
    \caption{
        Equal-time comparisons between regression-based and conventional Monte Carlo integration. We demonstrate the effectiveness of our approach with polynomials of order 1 to 3. The rendering involves 3D integrations (\textsc{Killeroos} and \textsc{Bathroom}). relMSE values are scaled by $10^3$. 
    }
    \label{fig:matrix}
\end{figure*}

\subsection{Direct matrix estimator}
It is well known that regression with a polynomial can be formulated as a solution to a linear system. 
The size of this linear system is proportional to the number of monomials $M$ and independent from the number of samples $N$, making it suitable for \ac{MC} integration. 

We can define a linear system by taking partial derivatives with respects to coefficients and setting all to zero 
\begin{align}
    \frac{\partial\REstimator(\theta)}{\partial c_q} = 0 \qquad \forall q \in \{0,1,\ldots,\numBasis\} \,.
\end{align}
Since $\REstimator(\theta)$ is the squared $\Lnorm{2}$ norm of the differences between $\IntegrandPSS(u_i)$ and $\CVfunctionPSS(u_i, \theta)$, the solution to this system is guaranteed to give a global minimum of $\REstimator(\theta)$.
In a matrix form, this linear system can be represented as
 \begin{align}
    \label{eq:least_squares_order_based}
    \begin{bmatrix}
        \BasisInnerProduct{0}{0} &  \cdots & \BasisInnerProduct{0}{\numBasis}\\
        \BasisInnerProduct{1}{0} &  \cdots & \BasisInnerProduct{1}{\numBasis}\\
        \vdots \\
        \BasisInnerProduct{\numBasis}{0} &  \cdots & \BasisInnerProduct{\numBasis}{\numBasis}\\
    \end{bmatrix}
    \begin{bmatrix}
        \Coefficient_0 \\
        \Coefficient_1 \\
        \vdots \\
        \Coefficient_\numBasis
    \end{bmatrix}
     =
    \begin{bmatrix}
        \Summation\IntegrandPSS(u_i)\Basis_0(u_i) \\
        \Summation\IntegrandPSS(u_i)\Basis_1(u_i) \\
        \vdots \\
        \Summation\IntegrandPSS(u_i)\Basis_\numBasis(u_i) 
    \end{bmatrix}
\end{align}
where the function $\BasisInnerProduct{\cdot}{\cdot}$ represents
\begin{align}
    \BasisInnerProduct{p}{q} = \Summation\Basis_p(u_i)\Basis_q(u_i)
\end{align}
which is precomputed for given bases $\Basis$. 
Updating the full matrix requires $M^2 + M$ floating point multiplication and summation if we ignore the cost of evaluating the monomials.

Since we need to evaluate the resulting $\CVfunctionPSS(u)$ at each of $u_i$ after regression, we chose to store all the $N$ samples first, solve the linear system, and then evaluate $\CVfunctionPSS(u_i)$ afterward. 
An iterative approach which does not require storing all the samples is possible under the matrix formulation~\cite{nakatsukasa2018approximate}, but we leave it as future work and use the simpler alternative to store all the samples. 

Note also that the size of the matrix stays the same regardless of the number of samples $N$.
We use complete orthogonal decomposition as a solver as it is fast for small matrices and does not require that the matrix is invertible (which may happen when $N$ is small). 
Algorithm~\ref{algo:DirectMatrixSolver} shows the details of this matrix-based approach.

\subsection{Gradient descent estimator}
\label{subsec:gradient_descent_estimator}
Since our formulation does not assume a specific approach for regression, we can use an alternative approach based on gradient descent. 
Rather than setting partial derivatives to zero to define a linear system, we use them as a gradient to iteratively update $\theta_k$ by starting from its initial guess $\theta_1$ and calculating 
\begin{align}
    \theta_{k+1} = \theta_{k} - \gamma \left(\cdots, \frac{\partial\REstimator(\theta_k)}{\partial c_{q,k}}, \cdots \right)
\end{align}
where $\gamma$ is a parameter which we set to $0.01$. 
Using \emph{stochastic} gradient descent, we can modify the above equation by observing that $u_i$ is a random sample used for Monte Carlo estimator $\REstimator(\theta)$ of $R(\theta)$ and use only \emph{one} sample for $k=1, \cdots, N$ as
\begin{align}
    \theta_{k+1} = \theta_{k} - \gamma \left(\cdots, \frac{\partial r(u_k, \theta_k) }{\partial c_{q,k}}. \cdots \right).
\end{align}
where we defined $r(u, \theta) = (\IntegrandPSS(u) - \CVfunctionPSS(u, \theta) )^2$ for simplicity.
This process can be repeated multiple times over the same set of samples to potentially reach a better solution for regression.

This gradient descent approach is quite general and can be used for more complex integrable functions than polynomials such as neural networks~\cite{Mueller:2020:NeuralCV}. 
In the case of polynomials, the computation of gradients is straightforward and only requires one evaluation of $\CVfunctionPSS(u)$ for each sample $u_i$ and the evaluations of the monomials.
We thus found that the gradient descent approach is generally less computationally costly than the matrix-based approach.
While more sophisticated algorithms may perform better, we have found that the above simple algorithm works well in our case.
Algorithm~\ref{algo:GradientDescentSolver} shows the details.

\SetKwInOut{Parameter}{Parameter}
\begin{algorithm}[t]
\Parameter{lr: learning rate, T: number of iterations}
\SetAlgoLined
 $f \leftarrow \{\}$; $u \leftarrow \{\}$\; 
 \For{$i \leftarrow 1$ \KwTo $N$}{
  $u_i \leftarrow$ random()\;
  $f_i \leftarrow \IntegrandPSS(u_i)$\;
 }
 $\theta_{\mbox{min}} \leftarrow 0^{M \times 1}$\;
 \For{$t \leftarrow 1$ \KwTo $T$}{
    $\theta_{\mbox{min}} \leftarrow$ gradient\_descent($\theta_{\mbox{min}}$, lr, $u_i$, $f_i$)\; 
 }
 \Return $G(\theta_{\mbox{min}}) + \frac{1}{N}(\sum_{i=1}^{N} f_i - \CVfunctionPSS(u_i, \theta_{\mbox{min}}))$\;
\caption{\label{algo:GradientDescentSolver}Gradient descent estimator.}
\end{algorithm}

\paragraph{Discussion}

The gradient descent approach is general and updating the coefficient is less costly compared to solving a matrix. 
The matrix approach can be used only when the residual is quadratic to the parameters, as in the case for polynomials. 
In more general cases where the residual is not a quadratic function of the parameters, one can still use the gradient descent approach. 
We advise to use the matrix for polynomials because it directly solves for the optimal solution.

At a low sample count, however, we found that the matrix approach can sometimes be numerically unstable as the matrix could be ill-conditioned.
We have observed this issue in particular for a higher-order polynomial in a high dimensional space, as the number of entries in the matrix will be large. 
In this case, the gradient descent approach might be better suited. 

Our regression is by no means perfect, and using more efficient regression techniques and implementation could lead to further performance improvement.

\begin{figure}[t!]
    \centering
    \includegraphics[width=\columnwidth]{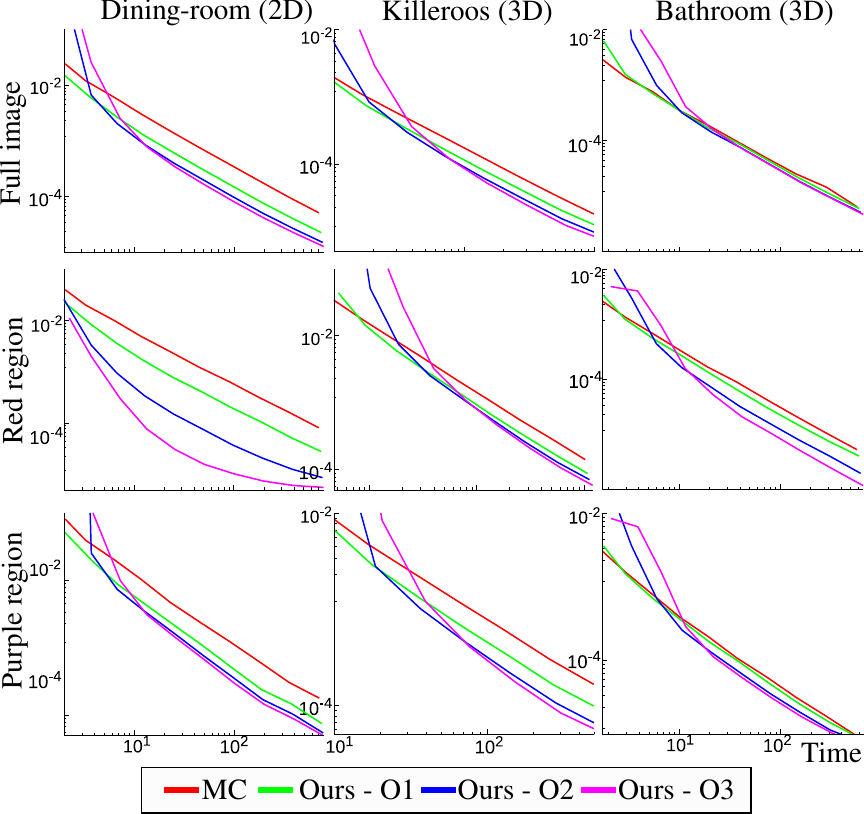}
    \caption{
        Convergence plot for equal-time comparisons in~\cref{fig:Teaser} and \cref{fig:matrix}. The metric is relMSE. 
        Our method converges faster than Monte Carlo estimation in all but has similar performance in the \textsc{Bathroom} scene.
    }
    \label{fig:convergence_plots}
\end{figure}

\section{Experiments}

Our implementation uses PBRT v3~\cite{Pharr:2016:Physically} as a rendering system and Eigen~\cite{eigen} as a solver for linear systems. 
All the results were generated on AMD EPYC 7702 64-Core or Intel Core i9-8950HK Processor using eight threads for each. For comparison, we compare our method against conventional \ac{MC} integration and \citet{crespo2020primaryspace} using the relative mean squared error (relMSE):

$
    \textstyle \nicefrac{1}{N} \sum_{i=1}^{N} \nicefrac{(I_i - R_i)^2}{(R_i^2 + 0.01)}, 
$

where $I$ and $R$ represent the rendered and the reference image, respectively. The reference image $R$ for each scene is computed with 65,536 samples per pixel. 

To handle a vector (i.e., RGB) integrand in rendering, we first estimate the RGB estimate $F_{\mathit{rgb}}$ by averaging RGB samples just as
$
 F_{\mathit{rgb}} \approx \langle F_{\mathit{rgb}} \rangle = \nicefrac{1}{N} \sum_{i=1}^{N} \IntegrandPSS_{\mathit{rgb}}(u_i)
$
where $\IntegrandPSS_{\mathit{rgb}}(u_i)$ is the integrand that returns an RGB sample. 
We then perform regression on the luminance value of the samples, resulting in the luminance valued $G$ as the analytical integral. The reconstruction of the RGB value based on our estimator is written as
$
    \langle F^*_{\mathit{rgb}} \rangle = (\nicefrac{\langle F_{\mathit{rgb}} \rangle }{y(\langle F_{\mathit{rgb}} \rangle)}) \CVEstimator
$
where $y$ is the luminance function and $\CVEstimator$ estimate the luminance value of $F_{\mathit{rgb}}$ based on our method. In other words, we correct the noisy luminance value of $ F_{\mathit{rgb}}$ by our estimate.

\subsection{Low-dimensional comparisons}

We show in~\Cref{fig:Teaser} and \Cref{fig:matrix} an equal-time comparison between the conventional \ac{MC} integration and our approach using polynomials of order 1 to 3 on three different scenes. 
The \textsc{Dining Room} scene (\Cref{fig:Teaser}) is composed of a single triangle light source for direct illumination, making it a 2-dimensional integration problem. Our method shows significant variance reduction compared to \ac{MC} integration as predicted. 
Even in the penumbra region where the reduction of error is small, our method is no worse than \ac{MC} integration.
The \textsc{Killeroos} scene (top row, \Cref{fig:matrix}) is composed of a single spherical light and motion blur, which is a 3-dimensional integration problem. Despite the fact that motion blur introduces more discontinuities in the integrand---which makes it difficult to approximate by polynomials of any order---our method can still reduce the variance in most of the regions.
This result demonstrates that it is possible to achieve variance reduction with model functions that only roughly approximate the underlying integrand. 

The \textsc{Bathroom} scene (bottom row, \Cref{fig:matrix}) has multiple mesh light sources picked randomly using a random number, making it also a 3-dimensional integration problem. Moreover, each light in this scene is a mesh light with several triangles making discontinuities in the $2$D area light parameterization. 
For this scene, the improvement in our method is limited but no worse than \ac{MC} integration. 

\Cref{fig:convergence_plots} shows the error convergence plots for the scenes described above. 
As predicted by our theory, our method consistently outperforms \ac{MC} integration once the solution to regression is almost converged. 
At a fewer sample count than 10 spp, our method with higher order polynomials (order two and three) may perform worse than Monte Carlo integration due to the error in regression.  
This is because the matrix tends to be ill-conditioned due to noise in the sampling. 
Even though a linear model function (O1) is only slightly more complex than a constant model function (\ac{MC}), our method with O1 significantly outperforms \ac{MC} integration.
We also observe that the maximum possible reduction in variance is limited by the complexity of the model function. There is thus a trade-off between the maximum reduction of variance and its stability at a few sample count, when one chooses a model function. The order or number of basis also seems to impact the ovearhead.

\begin{figure}[t!]
    \centering

\newcommand{\HighDimExempleTemplate}[1]{
    \begin{scope}
        \clip (0,0) -- (4.5,0) -- (4.5,4.5) -- (0,4.5) -- cycle;
        \path[fill overzoom image=figures/#1.jpg] (0,0) rectangle (4.5cm,4.5cm);
        
    \end{scope}
}

\newcommand{\PlotLargeImage}[1]{
    \begin{scope}
        \clip (0.7,0) -- (3.8,0) -- (3.8,3.2) -- (0.7,3.2) -- cycle;
        \path[fill overzoom image=figures/#1.jpg] (0,0) rectangle (4.5cm,3.2cm);
        \draw[draw=black,thick] (0.8,2.1) -- (0.8,2.7);
        \draw[draw=black,thick] (0.8,2.1) -- (1.4,2.1);
        \draw[draw=black,thick] (0.8,2.7) -- (1.4,2.7);
        \draw[draw=black,thick] (1.4,2.1) -- (1.4,2.7);
    \end{scope}
}

\newcommand{\PlotSplit}[3]{
    \begin{scope}
        \clip (0,0) -- (2.25,0) -- (0.0,4.5) -- cycle;
        \path[fill overzoom image=figures/#1] (0,0) rectangle (4.5cm,4.5cm);
    \end{scope}
    \begin{scope}
        \clip (2.25,0) -- (0.0,4.5) -- (2.25,4.5) -- (4.5,0) -- cycle;
        \path[fill overzoom image=figures/#2] (0,0) rectangle (4.5cm,4.5cm);
    \end{scope}
    \begin{scope}
        \clip (2.25,4.5) -- (4.5,0) -- (4.5,4.5) -- cycle;
        \path[fill overzoom image=figures/#3] (0,0) rectangle (4.5cm,4.5cm);
    \end{scope}
    \draw[draw=black,thick] (2.25,0) -- (0.0,4.5);
    \draw[draw=black,thick] (2.25,4.5) -- (4.5,0.0);
    \draw[draw=black,thick] (3.0,2.3) -- ((4.0,2.3) -- ((4.0,3.3) -- ((3.0,3.3) -- cycle;
}

\newcommand{\TestMethodTemplate}[2]{
    \begin{scope}
        \clip (0,0) -- (4.5,0) -- (0.0,4.5) -- cycle;
        \path[fill overzoom image=figures/#2] (0,0) rectangle (4.5,4.5);
    \end{scope}
    \begin{scope}
        \clip (0.0,4.5) -- (4.5,0) -- (4.5,4.5) -- cycle;
        \path[fill overzoom image=figures/#1] (0,0) rectangle (4.5,4.5);
    \end{scope}
    \draw[draw=black,thick] (4.5,0) -- (0.0,4.5);
}

\small
\hspace*{-2mm}
\setlength\tabcolsep{1pt}
\begin{tabularx}{\linewidth}{ccccc}

~ & Balance & + ours & Optimal & + ours\\
\multirow{3}{*}[0.38in]{\begin{tikzpicture}[scale=0.93]
    \PlotLargeImage{OptiMIS_book_64spp_direct/ref}
\end{tikzpicture}}
&
\begin{tikzpicture}[scale=0.27]
    \HighDimExempleTemplate{OptiMIS_book_64spp_direct/path_mis_balance_cvls_poly_o0_b2}
\end{tikzpicture}
&
\begin{tikzpicture}[scale=0.27]
    \HighDimExempleTemplate{OptiMIS_book_64spp_direct/path_mis_balance_cvls_poly_o2_b2}
\end{tikzpicture}
&
\begin{tikzpicture}[scale=0.27]
    \HighDimExempleTemplate{OptiMIS_book_64spp_direct/path_mis_direct_cvls_poly_o0_b2}
\end{tikzpicture}
&
\begin{tikzpicture}[scale=0.27]
    \HighDimExempleTemplate{OptiMIS_book_64spp_direct/path_mis_direct_cvls_poly_o2_b2}
\end{tikzpicture}
\\
&
\begin{tikzpicture}[scale=0.27]
    \HighDimExempleTemplate{OptiMIS_book_64spp_direct/path_mis_balance_cvls_poly_o0_mrse_b2}
\end{tikzpicture}
&
\begin{tikzpicture}[scale=0.27]
    \HighDimExempleTemplate{OptiMIS_book_64spp_direct/path_mis_balance_cvls_poly_o2_mrse_b2}
\end{tikzpicture}
&
\begin{tikzpicture}[scale=0.27]
    \HighDimExempleTemplate{OptiMIS_book_64spp_direct/path_mis_direct_cvls_poly_o0_mrse_b2}
\end{tikzpicture}
&
\begin{tikzpicture}[scale=0.27]
    \HighDimExempleTemplate{OptiMIS_book_64spp_direct/path_mis_direct_cvls_poly_o2_mrse_b2}
\end{tikzpicture}
\\
& 2.92E-3 & 1.96E-3 & 2.74E-3 & 1.86E-3
\\ 
& (a) & (b) & (c) & (d)

\end{tabularx}
    \caption{
       Equal sample comparison of two MIS techniques (BRDF+light sampling) is performed using balance heuristics \cite{Veach:1997:Robust} and optimal weights \cite{Kondapaneni:2019:Optimal}. The combination with our method uses an order-2 polynomial basis. 
       (a-d) Error maps and the relMSE values marked at the bottom shows improvement brought by our method.
    }
    \label{fig:mis_exemple}
\end{figure}
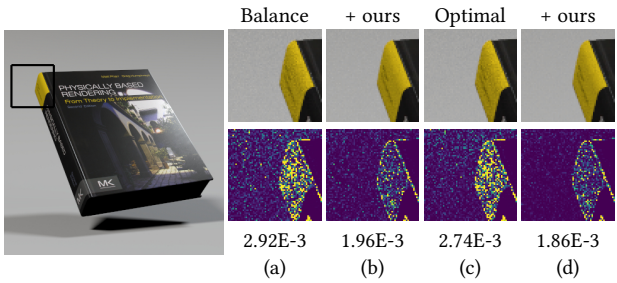

\subsection{Combination with multiple importance sampling}

In \Cref{fig:mis_exemple}, we show that our method can be used in conjunction with MIS~\cite{Veach:1995:Optimally}. From different sampling strategies available during \ac{MIS}, we simply perform the regression on weighted samples from each sampling strategy.

\Cref{fig:mis_exemple} shows the results using 64 samples per pixel of our method with two different importance sampling techniques: BRDF and light sampling. We compare the balance heuristic \citep{Veach:1997:Robust} and the optimal weights \citep{Kondapaneni:2019:Optimal}. 
Our method improves the quality of the result for both weighting schemes compared to Monte Carlo counterpart. 
We also found that the combination of our method with optimal weight~\citep{Kondapaneni:2019:Optimal} keeps the strength of these two techniques, leading to the best variance reduction.
The error reduction brought by the optimal weights allows a closer regression to the integrand leading to a greater error reduction. 
This result indicates that our method can take advantage of orthogonal variance reduction methods that simplify the shape of the integrand in primary sample space.

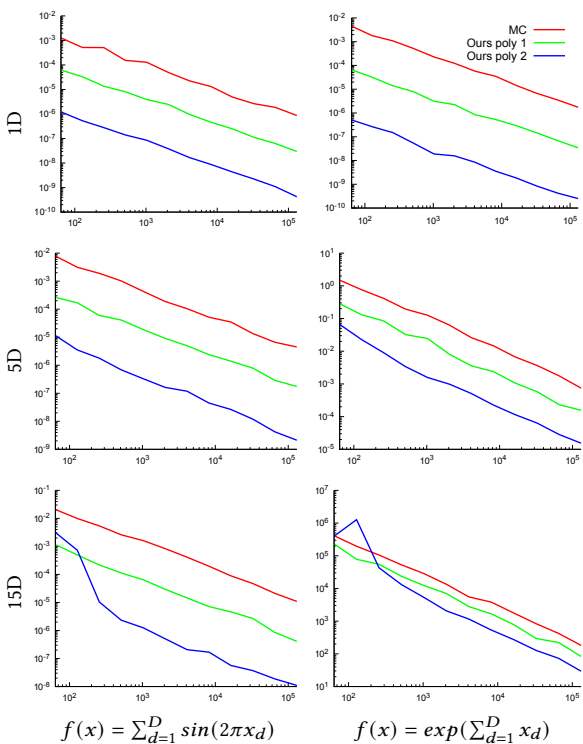
\begin{figure}[t!]
    \centering
    \newcommand{\ToyHightDim}[1]{
    \begin{scope}
        \clip (0,0) -- (4.5,0) -- (4.5,3.8) -- (0,3.8) -- cycle;
        \path[fill overzoom image=figures/#1.pdf] (0,0) rectangle (4.5cm,3.8cm);
    \end{scope}
}
\small
\hspace*{-2mm}
\setlength\tabcolsep{2pt}
\begin{tabularx}{\linewidth}{ccc}
\rotatebox{90}{\hspace{13mm}1D}
&
\begin{tikzpicture}[scale=0.8]
    \ToyHightDim{Toy_high_dim/graph_sum_sin_1D}
\end{tikzpicture}
&
\begin{tikzpicture}[scale=0.8]
    \ToyHightDim{Toy_high_dim/graph_exp_1D}
\end{tikzpicture}
\\
\rotatebox{90}{\hspace{13mm}5D}
&
\begin{tikzpicture}[scale=0.8]
    \ToyHightDim{Toy_high_dim/graph_sum_sin_5D}
\end{tikzpicture}
&
\begin{tikzpicture}[scale=0.8]
    \ToyHightDim{Toy_high_dim/graph_exp_5D}
\end{tikzpicture}
\\
\rotatebox{90}{\hspace{13mm}15D}
&
\begin{tikzpicture}[scale=0.8]
    \ToyHightDim{Toy_high_dim/graph_sum_sin_15D}
\end{tikzpicture}
&
\begin{tikzpicture}[scale=0.8]
    \ToyHightDim{Toy_high_dim/graph_exp_15D}
\end{tikzpicture}
\\
~&$f(x) = \sum_{d=1}^Dsin(2\pi x_d)$ & $f(x) = exp(\sum_{d=1}^D x_d)$
\end{tabularx}
    \caption{
    Convergence plots comparing our method using polynomial bases with a classical Monte Carlo estimator at different dimensionality. We observe that performance stays unchanged if the function complexity does not depend on the integration dimensionality (left column). Otherwise, our method experiences a slight decrease of improvement with the dimensionality but is still clearly advantageous compared to the Monte Carlo estimator (right column).
    }
    \label{fig:toy_example_high_dim}
\end{figure}

\begin{figure}[t!]
    \centering

\newcommand{\PlotSingleImage}[1]{
    \begin{scope}
        \clip (0,0) -- (4.5,0) -- (4.5,4.5) -- (0,4.5) -- cycle;
        \path[fill overzoom image=figures/#1.jpg] (0,0) rectangle (4.5cm,4.5cm);
        
    \end{scope}
}

\newcommand{\PlotRefImage}[1]{
    \begin{scope}
        \clip (0,0) -- (4.5,0) -- (4.5,4.5) -- (0,4.5) -- cycle;
        \path[fill overzoom image=figures/#1.jpg] (0,0) rectangle (4.5cm,4.5cm);
        \draw[draw=red,thick] (2.900390625,3.177) -- (2.900390625,2.0565);
        \draw[draw=red,thick] (2.900390625,3.177) -- (4.025390625,3.177);
        \draw[draw=red,thick] (2.900390625,2.0565) -- (4.025390625,2.0565);
        \draw[draw=red,thick] (4.025390625,3.177) -- (4.025390625,2.0565);
    \end{scope}
}
\small
\hspace*{-2mm}
\setlength\tabcolsep{2pt}
\begin{tabularx}{\linewidth}{ccc}
~ & MC & Ours\\
\multirow{3}{*}[0.59in]{\begin{tikzpicture}[scale=0.9]
    \PlotRefImage{Duck_cornell_path_3bounces_256spp/ref}
    \end{tikzpicture}}
&
\begin{tikzpicture}[scale=0.39]
    \PlotSingleImage{Duck_cornell_path_3bounces_256spp/cvls_poly_o0_b1}
\end{tikzpicture}
&
\begin{tikzpicture}[scale=0.39]
    \PlotSingleImage{Duck_cornell_path_3bounces_256spp/cvls_poly_o2_b1}
\end{tikzpicture}
\\
&
\begin{tikzpicture}[scale=0.39]
    \PlotSingleImage{Duck_cornell_path_3bounces_256spp/cvls_poly_o0_mrse_b1}
\end{tikzpicture}
&
\begin{tikzpicture}[scale=0.39]
    \PlotSingleImage{Duck_cornell_path_3bounces_256spp/cvls_poly_o2_mrse_b1}
\end{tikzpicture}
\\
&
 3.34E-3 & 2.39E-3
\end{tabularx}
    \caption{
        Equal sample comparison between classical Monte Carlo and our method using an \Order{2} polynomial basis. 
        In this \textsc{duck-cornell} scene, only the upper part is directly illuminated by the light source. 
        The lower part has indirect-illumination. The error map (relMSE) shows clear improvements using our method. relMSE values marked at the bottom.        
    }
    \label{fig:high_dim_exemple}
\end{figure}
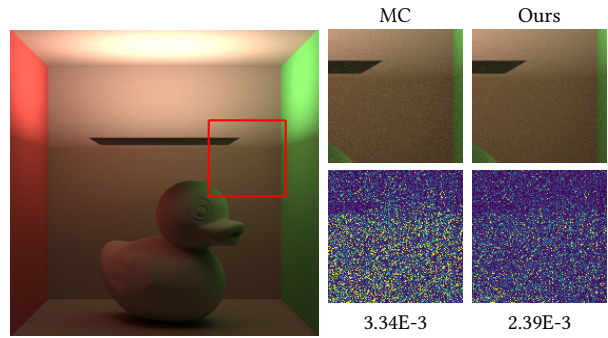

\begin{figure}[t!]
    \centering

\newcommand{\ComparisonBasisTemplate}[1]{
    \begin{scope}
        \clip (0,0) -- (6.0,0) -- (6.0,4.5) -- (0,4.5) -- cycle;
        \path[fill overzoom image=#1.jpg] (0,0) rectangle (6.0cm,4.5cm);
    \end{scope}
}
\newcommand{\PlotVertical}[1]{
    \begin{scope}
        \clip (0,0.5) -- (4.0,0.5) -- (4.0,4.5) -- (0,4.5) -- cycle;
        \path[fill overzoom image=#1.jpg] (0,0) rectangle (4.0cm,4.5cm);
        \draw[draw=red,thick] (2.94,2.17) -- (2.94,1.42);
        \draw[draw=red,thick] (2.94,2.17) -- (3.992,2.17);
        \draw[draw=red,thick] (2.94,1.42) -- (3.992,1.42);
        \draw[draw=red,thick] (3.992,2.17) -- (3.992,1.42);
    \end{scope}
}
\newcommand{\PlotCrop}[1]{
    \begin{scope}
        \clip (0,0) -- (5,0) -- (5,3.5) -- (0,3.5) -- cycle;
        \path[fill overzoom image=#1.jpg] (0,0) rectangle (5.0cm,3.5cm);
    \end{scope}
}

\small
\hspace*{-2mm}
\setlength\tabcolsep{1pt}
\begin{tabularx}{\linewidth}{ccccc}
~ &Reference & MC & Ours\\
\multirow{4}{*}[0.315in]{\begin{tikzpicture}[scale=0.84]
    \PlotVertical{figures/error_over_bouces_working_desk_2048spp/b1/cvls_poly_o0}
\end{tikzpicture}}

&
\begin{tikzpicture}[scale=0.3]
    \PlotCrop{figures/error_over_bouces_working_desk_2048spp/b1/ref_b1}
\end{tikzpicture}
&
\begin{tikzpicture}[scale=0.3]
    \PlotCrop{figures/error_over_bouces_working_desk_2048spp/b1/cvls_poly_o0_mrse_b1}
\end{tikzpicture}
& 
\begin{tikzpicture}[scale=0.3]
    \PlotCrop{figures/error_over_bouces_working_desk_2048spp/b1/cvls_poly_o2_mrse_b1}
\end{tikzpicture}
& \rotatebox{90}{\footnotesize 1-bounce}
\\
&
\begin{tikzpicture}[scale=0.3]
    \PlotCrop{figures/error_over_bouces_working_desk_2048spp/b2/ref_b1}
\end{tikzpicture}
&
\begin{tikzpicture}[scale=0.3]
    \PlotCrop{figures/error_over_bouces_working_desk_2048spp/b2/cvls_poly_o0_mrse_b1}
\end{tikzpicture}
&
\begin{tikzpicture}[scale=0.3]
    \PlotCrop{figures/error_over_bouces_working_desk_2048spp/b2/cvls_poly_o2_mrse_b1}
\end{tikzpicture}
& \rotatebox{90}{\footnotesize 2-bounces}
\\
&
\begin{tikzpicture}[scale=0.3]
    \PlotCrop{figures/error_over_bouces_working_desk_2048spp/b3/ref_b1}
\end{tikzpicture}
&
\begin{tikzpicture}[scale=0.3]
    \PlotCrop{figures/error_over_bouces_working_desk_2048spp/b3/cvls_poly_o0_mrse_b1}
\end{tikzpicture}
&
\begin{tikzpicture}[scale=0.3]
    \PlotCrop{figures/error_over_bouces_working_desk_2048spp/b3/cvls_poly_o2_mrse_b1}
\end{tikzpicture}
& \rotatebox{90}{\footnotesize 3-bounces}
\end{tabularx}
    \caption{
        We study the impact of our method on the function complexity by increasing the path length: 1, 2 and 3 bounces between the camera and the light source. The dimensionality of the respective path length is 3D, 5D and 7D. At equal sample count, the improvements of our method are directly affected by the complexity of the integrand.
    } 
    \label{fig:error_over_bouces}
\end{figure}
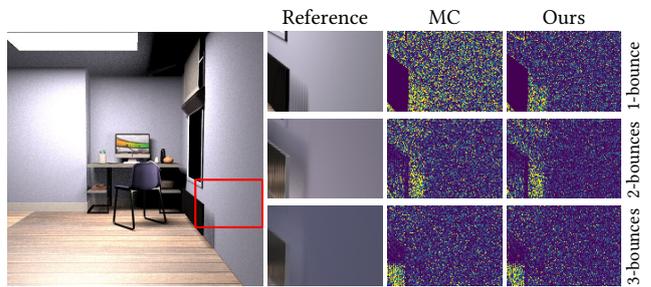

\begin{figure*}[t!]
    \centering
    \input{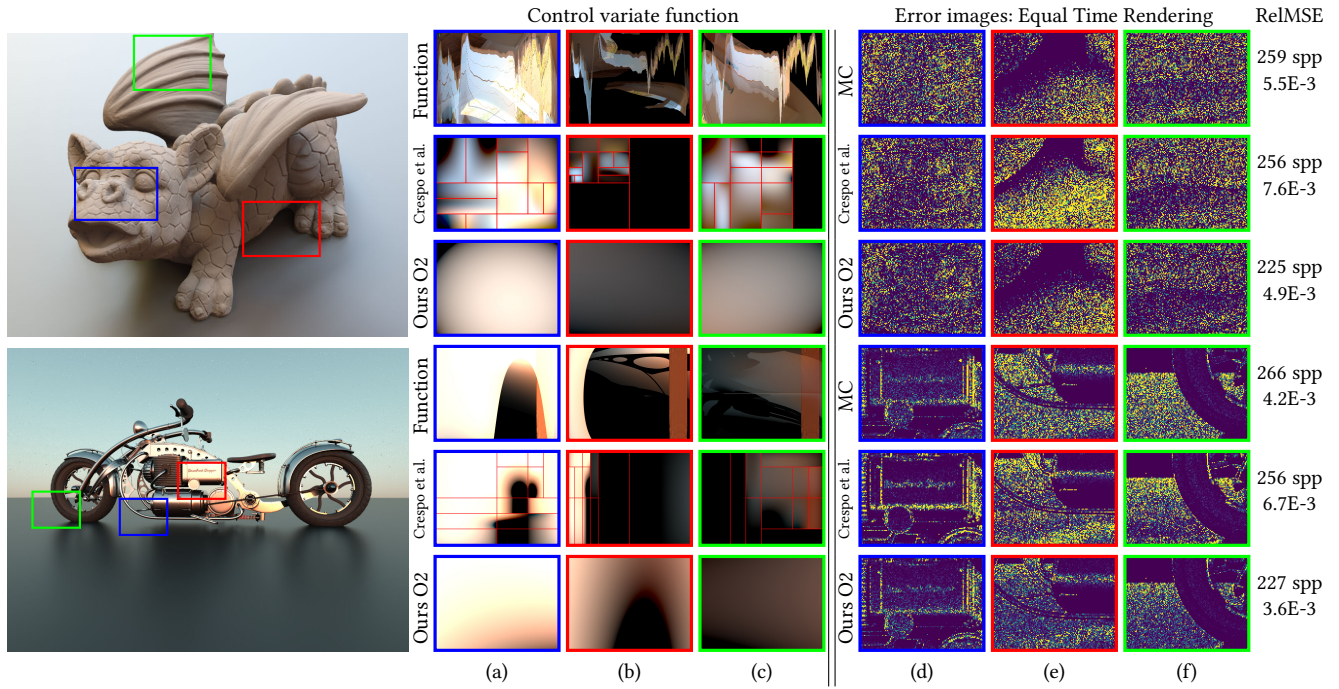}
    \caption{
       Equal-time comparison between our method and \citet{crespo2020primaryspace} where each pixel is considered independently. 
       Columns (a)-(c) show the control variate function for a pixel centered at each inset.  
       Columns (d)-(f) show the error images for each inset. Following \citeauthor{crespo2020primaryspace}, we ensure $30\%$ of the samples are used for building the control variate. 
       For both the \textsc{dragon} and the \textsc{chopper-titan} scene, our method show relatively less error ((d)-(f)). 
       \citeauthor{crespo2020primaryspace} is efficient in some cases. For \textsc{dragon} scene, column (e), \citeauthor{crespo2020primaryspace} gives worse error than the classical MC. 
       The relMSE value of our method is lower than \citeauthor{crespo2020primaryspace}.
       For better visualization, please see the \textsc{html} viewer in the supplemental.}
    \label{fig:crespoComp}
\end{figure*}

\begin{figure}[t!]
    \centering

\newcommand{\PlotSplitTwoImages}[2]{
    \begin{scope}
        \clip (0,0) -- (4.5,0) -- (4.5,3.5) -- cycle;
        \path[fill overzoom image=figures/#1] (0,0) rectangle (4.5,3.5);
    \end{scope}
    \begin{scope}
        \clip (0.0,0.0) -- (4.5,3.5) -- (0,3.5) -- cycle;
        \path[fill overzoom image=figures/#2] (0,0) rectangle (4.5,3.5);
    \end{scope}
    \draw[draw=white,thin] (0,0) -- (4.5,3.5);
}

\small
\hspace*{-5mm}
\setlength\tabcolsep{1pt}
\renewcommand{\arraystretch}{0.5} 
\begin{tabularx}{\linewidth}{c@{\;}c@{}}
    \adjustbox{valign=c}{\begin{tikzpicture}[scale=0.87]
        \PlotSplitTwoImages{VW-van_crespo_comp_256spp/Crespo_et_al}{VW-van_crespo_comp_256spp/Ours_Order_2_polynomial};
        \begin{scope}
            \filldraw[black,ultra thick] (3.1, 0.5) circle (0pt) node[anchor=north,rotate=0] 
            {\citet{crespo2020primaryspace} \tiny};
            \filldraw[black,ultra thick] (0.4, 3.5) circle (0pt) node[anchor=north,rotate=0] 
            {Ours \tiny};
            \filldraw[black,ultra thick] (-0.15, 0.90) circle (0pt) node[rotate=90, anchor=west] 
            {\textsc{VW-Van}};
        \end{scope}
    \end{tikzpicture}}
    &
    \adjustbox{valign=c}{\begin{tikzpicture}[scale=0.87]
        \PlotSplitTwoImages{VW-van_crespo_comp_256spp/Crespo_et_al-MRSE}{VW-van_crespo_comp_256spp/Ours_Order_2_polynomial-MRSE};
        \begin{scope}
            \filldraw[white,ultra thick] (3.1, 0.6) circle (0pt) node[anchor=north,rotate=0] 
            {relMSE: $2.064\times 10^{-3}$ \tiny};
            \filldraw[white,ultra thick] (1.4, 3.5) circle (0pt) node[anchor=north,rotate=0] 
            {relMSE: $1.237\times 10^{-3}$ \tiny};
        \end{scope}
    \end{tikzpicture}}
    \\
    \adjustbox{valign=c}{\begin{tikzpicture}[scale=0.87]
        \PlotSplitTwoImages{Teapot_crespo_comp_256spp/Crespo_et_al}{Teapot_crespo_comp_256spp/Ours_Order_2_polynomial};
         \filldraw[black,ultra thick] (-0.15, 0.90) circle (0pt) node[rotate=90, anchor=west] 
            {\textsc{Teapot}};
    \end{tikzpicture}}
    &
    \adjustbox{valign=c}{\begin{tikzpicture}[scale=0.87]
        \PlotSplitTwoImages{Teapot_crespo_comp_256spp/Crespo_et_al-MRSE}{Teapot_crespo_comp_256spp/Ours_Order_2_polynomial-MRSE};
        \begin{scope}
            \filldraw[white,ultra thick] (3.5, 0.6) circle (0pt) node[anchor=north,rotate=0] 
            {$6.174\times 10^{-3}$ \tiny};
            \filldraw[white,ultra thick] (1.0, 3.5) circle (0pt) node[anchor=north,rotate=0] 
            {$3.477\times 10^{-3}$ \tiny};
        \end{scope}
    \end{tikzpicture}}
    \\
    \adjustbox{valign=c}{\begin{tikzpicture}[scale=0.87]
        \PlotSplitTwoImages{House_crespo_comp_256spp/Crespo_et_al}{House_crespo_comp_256spp/Ours_Order_2_polynomial};
        \filldraw[black,ultra thick] (-0.15, 0.90) circle (0pt) node[rotate=90, anchor=west] 
            {\textsc{House}};
    \end{tikzpicture}}
    &
    \adjustbox{valign=c}{\begin{tikzpicture}[scale=0.87]
        \PlotSplitTwoImages{House_crespo_comp_256spp/Crespo_et_al-MRSE}{House_crespo_comp_256spp/Ours_Order_2_polynomial-MRSE};
        \begin{scope}
            \filldraw[white,ultra thick] (3.5, 0.6) circle (0pt) node[anchor=north,rotate=0] 
            {$2.412\times 10^{-3}$ \tiny};
            \filldraw[white,ultra thick] (1.0, 3.5) circle (0pt) node[anchor=north,rotate=0] 
            {$2.572\times 10^{-3}$ \tiny};
        \end{scope}
    \end{tikzpicture}}
\end{tabularx}
    \caption{
       Equal-sample comparison of our method using an $\Order{2}$ polynomial function basis with \citet{crespo2020primaryspace} method.}
    \label{fig:crespoComp2}
\end{figure}
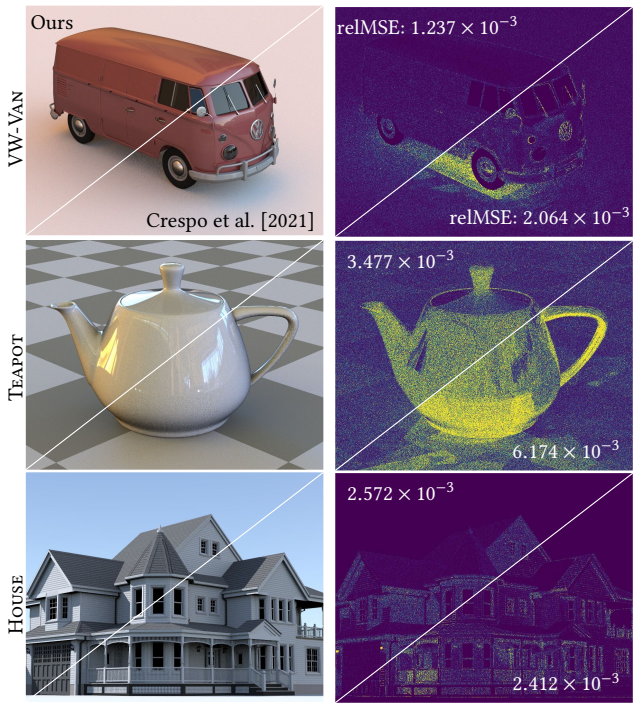

\subsection{High-dimensional comparisons}

Our method directly extends to higher dimensions. 

We start the numerical analysis with simple analytic integrands. In \cref{fig:toy_example_high_dim}, we show the convergence curves over 1D, 5D and 15D space for two integrands: (a) sum of sinusoids and (b) a multi-dimensional exponential function. 
In both cases, our method shows significant improvements in variance reduction. Both \Order{1} and \Order{2} polynomials gives significant improvements. However, at low sample count (up to about 100 samples) in 15D, \Order{2} polynomial performs poorly. This is because with increasing polynomial order, in higher dimensions, more samples are needed to better approximate the control variate.

We apply our method to a path tracing rendering. 
\Cref{fig:high_dim_exemple} shows the results of our method for a \textsc{duck-cornell} scene rendered in path tracing with 3 bounces. 
This corresponds to a 15D problem in our algorithm: a 3D random number for each next-event estimation and a 2D random number for the next-direction sampling at each bounce. 
Improvements are visible in both directly- and indirectly-illumniated regions. 
However, the improvements in the indirectly-lit regions are less due to the strong variations of the function. These variations come from the light making several bounces, which makes the control variate function less accurate than in the case of direct illumination. 
One can of course consider having a more complex model function, but it will put an additional computational overhead on our method.

\Cref{fig:error_over_bouces} shows the improvement due to our method on illumination coming from different number of bounces.
Our method provides the largest improvement on direct illumination (the top row), and the improvement generally decreases with more bounces (the middle and the bottom rows).
As we discussed above, it is not directly due to the increased number of dimensions for higher number of bounces.
We speculate that it is rather specific to the fact that the integrand in the primary sample space becomes more and more complex for higher number of bounces. 
Similar observations have been made in prior work that attempt to approximate the integrand in the primary sample space~\cite{guo2018primary,zheng2019learning,crespo2020primaryspace}, so we believe that it is not specific to our regression approach. 
We note that our method is still not performing worse than \ac{MC} integration even for higher number of bounces.

\subsection{Comparison to \citet{crespo2020primaryspace}}
%
\citet{crespo2020primaryspace} proposed two adaptive control variate techniques using quadratures: one based on a quadrature per pixel and one with a global quadrature for the whole image. As we also treat each pixel independently, we have chosen to compare our method to the former method, which has been shown to be competitive. 
Specifically, \citet{crespo2020primaryspace} construct an adaptive piecewise-polynomial control variate (per pixel) in the primary sample space by recursively splitting the space based on an error heuristic built upon nested quadrature rules and regions' volume. 
We leave the possibility of extending our approach over a block of pixels as future work.

\Cref{fig:crespoComp} provides an equal-time comparison of \citeauthor{crespo2020primaryspace}'s approach with our $\Order{2}$ polynomial basis method for a $2$D direct illumination integration in two scenes. The \textsc{dragon} scene consist of a simple object lit by a high-frequcency environment map. The \textsc{chopper-titan} scene contains glossy complex objects lit by a low-frequency environment map. For both scenes, we show the reference control variate function, the control variate functions constructed by \citeauthor{crespo2020primaryspace} and our method (columns (a)-(c)). We also shows on columns (d)-(f) the error images of MC and these different technique.

For functions shown in column (a), \citeauthor{crespo2020primaryspace}'s method can produce very accurate function approximations, resulting in error reduction compared to MC method (d). 
However, in columns (b, c), their adaptive control variate construction misses some critical function regions, resulting in a poor approximation. Moreover, stratifying the samples across regions with importance sampling increases this problem, resulting in higher error than MC (columns e,f).
In comparison, our method uses a crude approximation of control variate via a polynomial function but consistently outperforms the classical MC and is more robust compared to \citeauthor{crespo2020primaryspace}'s method.

\Cref{fig:crespoComp2} shows an equal-sample comparison between \citeauthor{crespo2020primaryspace}'s method and ours. 
For the \textsc{vw-van} and the \textsc{teapot} scenes, our method significantly outperforms \citeauthor{crespo2020primaryspace}'s in difficult regions such as the bottom of the van and the reflections on the bottom side of the teapot. However, for the \textsc{house} scene, \citeauthor{crespo2020primaryspace}'s method performs better especially on the house's facades. 
Despite such, it is noteworthy that the error reduction by our method is more uniform across the scenes. 
\citeauthor{crespo2020primaryspace}'s method, on the other hand, shows significant improvement upon classical MC in some regions, e.g., directly illuminated regions of the \textsc{vw-van} and the \textsc{teapot}, but performs worse than MC in some other regions.

\begin{figure}[t!]
    \centering

\newcommand{\PlotSingleImage}[1]{
    \begin{scope}
        \clip (0,0) -- (4.5,0) -- (4.5,4.5) -- (0,4.5) -- cycle;
        \path[fill overzoom image=figures/#1.jpg] (0,0) rectangle (4.5cm,4.5cm);
        
    \end{scope}
}

\newcommand{\PlotRefImage}[1]{
    \begin{scope}
        \clip (0,0) -- (4.5,0) -- (4.5,4.5) -- (0,4.5) -- cycle;
        \path[fill overzoom image=figures/#1.jpg] (0,0) rectangle (4.5cm,4.5cm);
        \draw[draw=red,thick] (0.400390625,1.677) -- (0.400390625,0.0565);
        \draw[draw=red,thick] (0.400390625,1.677) -- (2.025390625,1.677);
        \draw[draw=red,thick] (0.400390625,0.0565) -- (2.025390625,0.0565);
        \draw[draw=red,thick] (2.025390625,1.677) -- (2.025390625,0.0565);
    \end{scope}
}
\small
\hspace*{-2mm}
\setlength\tabcolsep{2pt}
\begin{tabularx}{\linewidth}{ccccc}
 & 
 MC 
 & 
 \multicolumn{2}{c}{
 \citet{crespo2020primaryspace}
 }
 & Ours

\\
\begin{tikzpicture}[scale=0.34]
    \PlotSingleImage{Crespo_no_IS/ref_b4}
\end{tikzpicture}
&
\begin{tikzpicture}[scale=0.34]
    \PlotSingleImage{Crespo_no_IS/cvls_poly_o0_mrse_b4}
\end{tikzpicture}
&
\begin{tikzpicture}[scale=0.34]
    \PlotSingleImage{Crespo_no_IS/crespo_pixel_noIS_mrse_b4}
\end{tikzpicture}
&
\begin{tikzpicture}[scale=0.34]
    \PlotSingleImage{Crespo_no_IS/crespo_pixel_mrse_b4}
\end{tikzpicture}
&
\begin{tikzpicture}[scale=0.34]
    \PlotSingleImage{Crespo_no_IS/cvls_poly_o2_mrse_b4}
\end{tikzpicture}
\\

\begin{tikzpicture}[scale=0.34]
    \PlotSingleImage{Crespo_no_IS/ref_b3}
\end{tikzpicture}
&
\begin{tikzpicture}[scale=0.34]
    \PlotSingleImage{Crespo_no_IS/cvls_poly_o0_mrse_b3}
\end{tikzpicture}
&
\begin{tikzpicture}[scale=0.34]
    \PlotSingleImage{Crespo_no_IS/crespo_pixel_noIS_mrse_b3}
\end{tikzpicture}
&
\begin{tikzpicture}[scale=0.34]
    \PlotSingleImage{Crespo_no_IS/crespo_pixel_mrse_b3}
\end{tikzpicture}
&
\begin{tikzpicture}[scale=0.34]
    \PlotSingleImage{Crespo_no_IS/cvls_poly_o2_mrse_b3}
\end{tikzpicture}
\\
 &  & w/o IS & w/ IS & 
\end{tabularx}
    \caption{
        Importance sampling (IS) performed in \citeauthor{crespo2020primaryspace}'s approach can increase the variance wrt classical MC due to bad IS strategy. 
    }
    \label{fig:crespo_no_IS}
\end{figure}
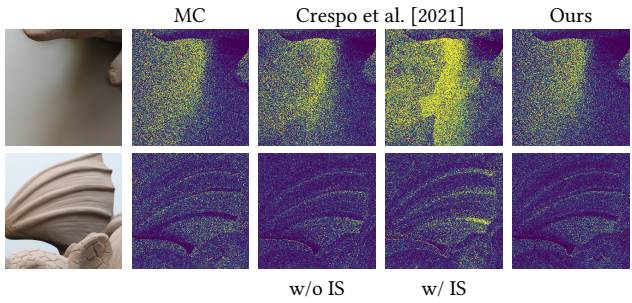

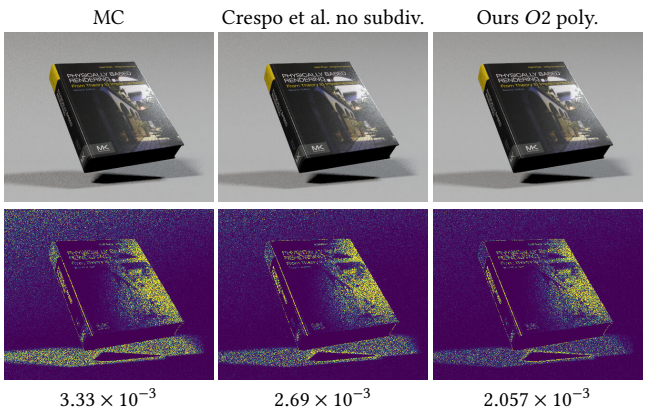
\begin{figure}[t!]
    \centering

\newcommand{\HighDimExempleTemplate}[1]{
    \begin{scope}
        \clip (0,0) -- (5.5,0) -- (5.5,4.5) -- (0,4.5) -- cycle;
        \path[fill overzoom image=figures/#1.jpg] (0,0) rectangle (5.5cm,4.5cm);
        
    \end{scope}
}

\newcommand{\PlotLargeImage}[1]{
    \begin{scope}
        \clip (0,0) -- (4.4,0) -- (4.4,3.2) -- (0,3.2) -- cycle;
        \path[fill overzoom image=figures/#1.jpg] (0,0) rectangle (4.4cm,3.2cm);
        
    \end{scope}
}

\newcommand{\PlotSplit}[3]{
    \begin{scope}
        \clip (0,0) -- (2.25,0) -- (0.0,4.5) -- cycle;
        \path[fill overzoom image=figures/#1] (0,0) rectangle (4.5cm,4.5cm);
    \end{scope}
    \begin{scope}
        \clip (2.25,0) -- (0.0,4.5) -- (2.25,4.5) -- (4.5,0) -- cycle;
        \path[fill overzoom image=figures/#2] (0,0) rectangle (4.5cm,4.5cm);
    \end{scope}
    \begin{scope}
        \clip (2.25,4.5) -- (4.5,0) -- (4.5,4.5) -- cycle;
        \path[fill overzoom image=figures/#3] (0,0) rectangle (4.5cm,4.5cm);
    \end{scope}
    \draw[draw=black,thick] (2.25,0) -- (0.0,4.5);
    \draw[draw=black,thick] (2.25,4.5) -- (4.5,0.0);
    \draw[draw=black,thick] (3.0,2.3) -- ((4.0,2.3) -- ((4.0,3.3) -- ((3.0,3.3) -- cycle;
}

\newcommand{\TestMethodTemplate}[2]{
    \begin{scope}
        \clip (0,0) -- (4.5,0) -- (0.0,4.5) -- cycle;
        \path[fill overzoom image=figures/#2] (0,0) rectangle (4.5,4.5);
    \end{scope}
    \begin{scope}
        \clip (0.0,4.5) -- (4.5,0) -- (4.5,4.5) -- cycle;
        \path[fill overzoom image=figures/#1] (0,0) rectangle (4.5,4.5);
    \end{scope}
    \draw[draw=black,thick] (4.5,0) -- (0.0,4.5);
}

\small
\hspace*{-2mm}
\setlength\tabcolsep{1pt}
\begin{tabularx}{\linewidth}{ccc}
 MC & \citeauthor{crespo2020primaryspace} no subdiv. & Ours $O2$ poly.
\\
\begin{tikzpicture}[scale=0.5]
    \HighDimExempleTemplate{Crespo_no_subdiv/cvls_poly_o0}
\end{tikzpicture}
&
\begin{tikzpicture}[scale=0.5]
    \HighDimExempleTemplate{Crespo_no_subdiv/crespo_pixel_noSubdiv}
\end{tikzpicture}
&
\begin{tikzpicture}[scale=0.5]
    \HighDimExempleTemplate{Crespo_no_subdiv/cvls_poly_o2}
\end{tikzpicture}
\\
    \begin{tikzpicture}[scale=0.5]
        \HighDimExempleTemplate{Crespo_no_subdiv/cvls_poly_o0_mrse}
    \end{tikzpicture}
    &
    \begin{tikzpicture}[scale=0.5]
        \HighDimExempleTemplate{Crespo_no_subdiv/crespo_pixel_noSubdiv_mrse}
    \end{tikzpicture}
    &
    \begin{tikzpicture}[scale=0.5]
        \HighDimExempleTemplate{Crespo_no_subdiv/cvls_poly_o2_mrse}
    \end{tikzpicture}
\\
$3.33 \times 10^{-3}$ & $2.69 \times 10^{-3}$ & $2.057 \times 10^{-3}$

\end{tabularx}
    \caption{
        Without recursive space subdivision, \citeauthor{crespo2020primaryspace} appears to be similar to our approach, but it is not. Our approach still performs better. All methods are rendered with 64 spp. relMSE values are at the bottom.
    }
    \label{fig:crespo_no_Subdiv}
\end{figure}

\Cref{fig:crespo_no_IS} demonstrates the impact of importance sampling in \citeauthor{crespo2020primaryspace}'s method. 
We found that their importance sampling strategy can increase variance, which can be explained by closely investigating their adaptive subdivision scheme. 
In the first row of~\cref{fig:crespoComp}, \citeauthor{crespo2020primaryspace}'s control variate function misses important integrand features. They used this approximation for importance sampling as well which may increase variance.

\Cref{fig:crespo_no_Subdiv} shows another experiment with no adaptive subdivision in the method of ~\citeauthor{crespo2020primaryspace}. 
Without adaptive subdivision, their method falls back to a control variate estimator using a simple quadrature for a polynomial model function. 
This setup looks deceivingly equivalent to ours, but they are different.
%
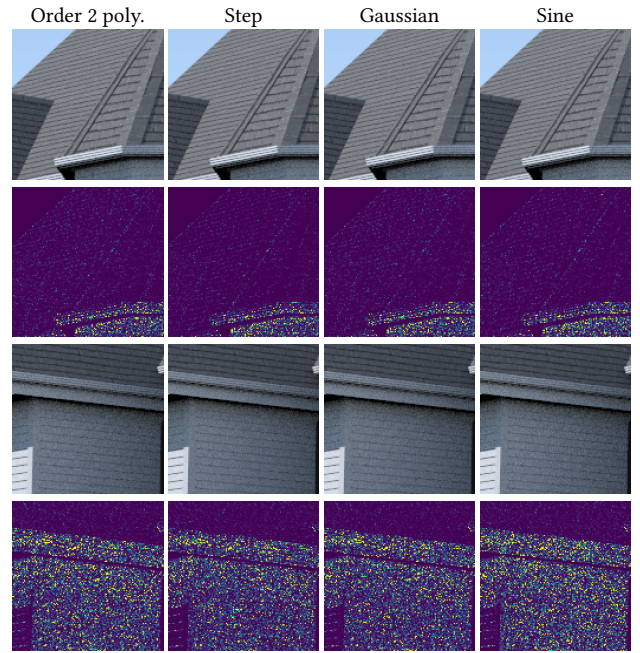
\begin{figure}[t!]
    \centering

\newcommand{\HighDimExempleTemplate}[1]{
    \begin{scope}
        \clip (0,0) -- (4.5,0) -- (4.5,4.5) -- (0,4.5) -- cycle;
        \path[fill overzoom image=figures/#1.jpg] (0,0) rectangle (4.5cm,4.5cm);
        
    \end{scope}
}

\newcommand{\PlotLargeImage}[1]{
    \begin{scope}
        \clip (0,0) -- (4.4,0) -- (4.4,3.2) -- (0,3.2) -- cycle;
        \path[fill overzoom image=figures/#1.jpg] (0,0) rectangle (4.4cm,3.2cm);
        
    \end{scope}
}

\newcommand{\PlotSplit}[3]{
    \begin{scope}
        \clip (0,0) -- (2.25,0) -- (0.0,4.5) -- cycle;
        \path[fill overzoom image=figures/#1] (0,0) rectangle (4.5cm,4.5cm);
    \end{scope}
    \begin{scope}
        \clip (2.25,0) -- (0.0,4.5) -- (2.25,4.5) -- (4.5,0) -- cycle;
        \path[fill overzoom image=figures/#2] (0,0) rectangle (4.5cm,4.5cm);
    \end{scope}
    \begin{scope}
        \clip (2.25,4.5) -- (4.5,0) -- (4.5,4.5) -- cycle;
        \path[fill overzoom image=figures/#3] (0,0) rectangle (4.5cm,4.5cm);
    \end{scope}
    \draw[draw=black,thick] (2.25,0) -- (0.0,4.5);
    \draw[draw=black,thick] (2.25,4.5) -- (4.5,0.0);
    \draw[draw=black,thick] (3.0,2.3) -- ((4.0,2.3) -- ((4.0,3.3) -- ((3.0,3.3) -- cycle;
}

\newcommand{\TestMethodTemplate}[2]{
    \begin{scope}
        \clip (0,0) -- (4.5,0) -- (0.0,4.5) -- cycle;
        \path[fill overzoom image=figures/#2] (0,0) rectangle (4.5,4.5);
    \end{scope}
    \begin{scope}
        \clip (0.0,4.5) -- (4.5,0) -- (4.5,4.5) -- cycle;
        \path[fill overzoom image=figures/#1] (0,0) rectangle (4.5,4.5);
    \end{scope}
    \draw[draw=black,thick] (4.5,0) -- (0.0,4.5);
}

\small
\hspace*{-2mm}
\setlength\tabcolsep{1pt}
\begin{tabularx}{\linewidth}{cccc}
 Order 2 poly. & Step & Gaussian & Sine
\\
\begin{tikzpicture}[scale=0.44]
    \HighDimExempleTemplate{basis_comparison/house_128/cvls_poly_o2_b1}
\end{tikzpicture}
&
\begin{tikzpicture}[scale=0.44]
    \HighDimExempleTemplate{basis_comparison/house_128/cvls_step_b1}
\end{tikzpicture}
&
\begin{tikzpicture}[scale=0.44]
    \HighDimExempleTemplate{basis_comparison/house_128/cvls_gaussian_b1}
\end{tikzpicture}
&
\begin{tikzpicture}[scale=0.44]
        \HighDimExempleTemplate{basis_comparison/house_128/cvls_sin_b1}
\end{tikzpicture}
\\
    \begin{tikzpicture}[scale=0.44]
        \HighDimExempleTemplate{basis_comparison/house_128/cvls_poly_o2_mrse_b1}
    \end{tikzpicture}
    &
    \begin{tikzpicture}[scale=0.44]
        \HighDimExempleTemplate{basis_comparison/house_128/cvls_step_mrse_b1}
    \end{tikzpicture}
    &
    \begin{tikzpicture}[scale=0.44]
        \HighDimExempleTemplate{basis_comparison/house_128/cvls_gaussian_mrse_b1}
    \end{tikzpicture}
    &
\begin{tikzpicture}[scale=0.44]
        \HighDimExempleTemplate{basis_comparison/house_128/cvls_sin_mrse_b1}
\end{tikzpicture}
    \\
    
    \begin{tikzpicture}[scale=0.44]
        \HighDimExempleTemplate{basis_comparison/house_128/cvls_poly_o2_b2}
    \end{tikzpicture}
    &
    \begin{tikzpicture}[scale=0.44]
        \HighDimExempleTemplate{basis_comparison/house_128/cvls_step_b2}
    \end{tikzpicture}
    &
    \begin{tikzpicture}[scale=0.44]
        \HighDimExempleTemplate{basis_comparison/house_128/cvls_gaussian_b2}
    \end{tikzpicture}
    &
    \begin{tikzpicture}[scale=0.44]
        \HighDimExempleTemplate{basis_comparison/house_128/cvls_sin_b2}
    \end{tikzpicture}
    \\
    \begin{tikzpicture}[scale=0.44]
        \HighDimExempleTemplate{basis_comparison/house_128/cvls_poly_o2_mrse_b2}
    \end{tikzpicture}
    &
    \begin{tikzpicture}[scale=0.44]
        \HighDimExempleTemplate{basis_comparison/house_128/cvls_step_mrse_b2}
    \end{tikzpicture}
    &
    \begin{tikzpicture}[scale=0.44]
        \HighDimExempleTemplate{basis_comparison/house_128/cvls_gaussian_mrse_b2}
    \end{tikzpicture}
    &
\begin{tikzpicture}[scale=0.44]
        \HighDimExempleTemplate{basis_comparison/house_128/cvls_sin_mrse_b2}
\end{tikzpicture}
    \\
\end{tabularx}
    \caption{We compare our method with different basis functions at equal sample count. Please refer to the supplemental material for more scenes and bases. 
    }
    \label{fig:comparison_basis}
\end{figure}

\citeauthor{crespo2020primaryspace}'s
method finds a polynomial via a quadrature rule with a fixed set of points within the domain for their control variates. 
Consequently, their control variate cannot take advantage of any more samples than being required for the quadrature rule without splitting. 
Our method, on the other hand, can take any number of samples to improve the least-squares fit without any splitting.
The least-squares regression further makes sure to reduce the variance bound as our formulation shows.
Both the error map and the relMSE values reported in~\cref{fig:crespo_no_Subdiv} confirm our observation.

In summary, due to the lack of robustness during the quadrature construction caused by their fixed sample locations and error heuristics,  \citeauthor{crespo2020primaryspace}'s method does not consistently outperform the classical MC overall.
In the case of inaccurate subdivisions of the primary sample space, their method might result in inefficient importance sampling, leading to higher errors than MC. 
By contrast, our method is provably never worse than the MC estimator despite that ours could result in rather coarse approximations of the function, limiting the error reduction in some regions. 
As future work, \citeauthor{crespo2020primaryspace}'s adaptive model for $\CVfunctionPSS$ can potentially be useful in our formulation, but developing a practical least-squares regression algorithm for piecewise polynomials remains challenging.

\subsection{Analysis}

\paragraph{Choice of basis functions}
We perform an experiment using different basis functions in the regression. 
The results are shown in~\cref{fig:comparison_basis}. 
We compare the performance of order 2 polynomial, step functions, Gaussian mixture and Sine functions. All these basis have been chosen to contain the same number of parameters for a fair comparison. 
The quality of the results obtained slightly varies according to the bases used and the scenes chosen.  
Other results for additional bases and scenes is available in the supplemental materials.
%
%
\begin{figure}[b!]
    \centering
    \includegraphics[width=\columnwidth]{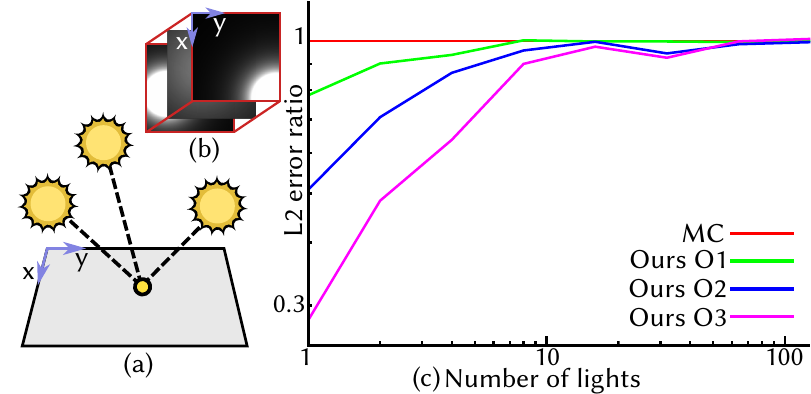}
    \caption{
    Comparison of integration error ratio with Monte Carlo as a function of the number of lights in the scene. Our method is used here with polynomial basis of order 1, 2 and 3. 
    The drawing on the left schematizes the integration problem. 
    }
    \label{fig:multi_loghts_experiment}
\end{figure}
%
%
\begin{figure*}[t!]
    \centering
    \includegraphics[width=\textwidth]{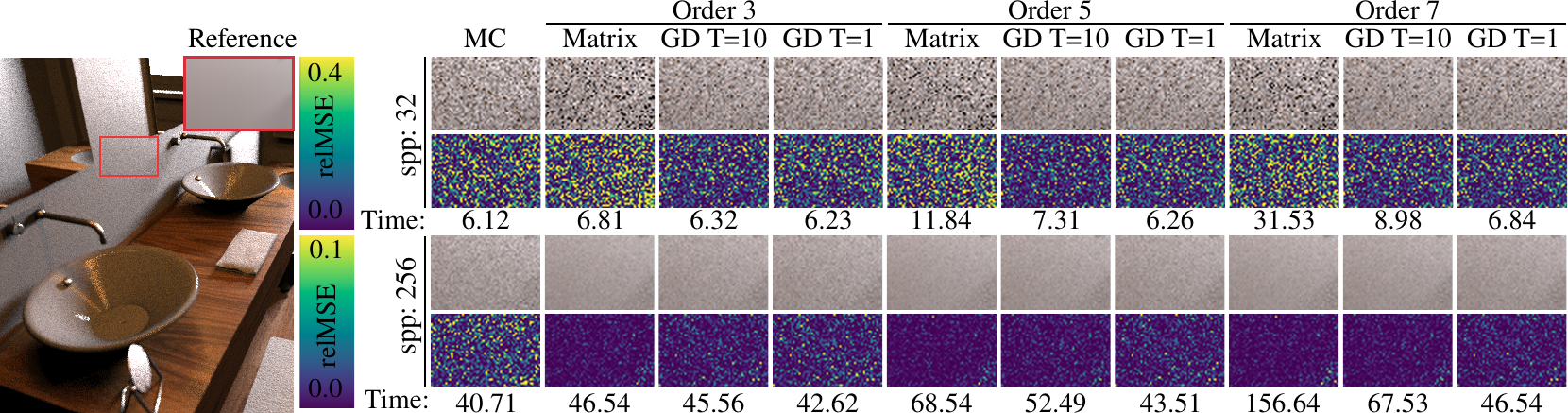}
    \caption{
        Regression using a direct matrix approach and stochastic gradient descent (SGD). At a low-sample count, SGD performs better. At a high-sample count, the direct matrix approach outperforms SGD. At high-order polynomial (order 7), the direct matrix approach has a very high overhead, making SGD an attractive alternative. Time is reported in seconds.
    }
    \label{fig:gd_experiment}
\end{figure*}

In general, while it is  challenging to define a function basis robust for all situations, our experiments show that polynomial basis works relatively well in many cases. As long as a constant function is included in least-squares regression, our method would be no worse than \ac{MC} integration, no matter, what basis functions are chosen. 

\paragraph{Impact of discontinuities on polynomial basis}


Polynomial bases have difficulty in representing functions with many discontinuities. 
In~\cref{fig:multi_loghts_experiment}, we study the evolution of the integration error for a fixed sample budget of 1024 samples when integrating the contribution of multiple light sources over a diffuse plane (\cref{fig:multi_loghts_experiment}.a). 
In this case, the number of discontinuities inside the $3$D integrand increases with the number of lights due to the random decision on the light selection (\cref{fig:multi_loghts_experiment}.b). 
We study the impact of using \Order{1}, \Order{2} and \Order{3} polynomials with our method against the conventional Monte Carlo estimator. 
The plot in~\cref{fig:multi_loghts_experiment}(c) demonstrates that at a low number of lights, our method significantly improves over the conventional Monte Carlo estimation. However, this improvement decreases with the increase in the number of discontinuities. With a large number of lights, the error produced by our method becomes equivalent to that of the Monte Carlo estimator, confirming that the polynomial basis behaves equivalently to a constant model function, \ie the conventional approach.

\paragraph{Computation time} 
We empirically found that the computation overhead of our method becomes significant for higher-order polynomial regression using the direct matrix approach. 
For example, our method with low-order polynomials has negligible overhead over \ac{MC} integration (\cref{fig:matrix}) but a polynomial of order $5$ can cause $68\%$ overhead in terms of running time. 
Similarly, higher-dimensional integration also requires significant overhead over \ac{MC} integration. 
However, we focus in demonstrating the versatility of our approach. 
Improving the regression computation is out of the scope for this work.
Following above observation, we also investigate another approach for regression, such as gradient descent, which can be a better option for higher-order polynomials. 
The gradient descent approach scales well thanks to its lower computational complexity ($O(M)$ vs $O(M^2)$) where $M$ is the number of monomials.

\paragraph{Matrix vs. Gradient descent solver}
~\Cref{fig:gd_experiment} shows the comparison between the gradient descent and the direct-matrix solver. The results are demonstrated with 32spp and 256spp on a \textsc{bathroom} scene taken from PBRT. We use three different polynomials of order 2, 5 and 7. 
We compare the direct matrix approach based on complete orthogonal decomposition and stochastic gradient descent estimator (SGD) to solve the polynomial regression. 
At a low sample count, SGD performs better than the direct matrix approach as the matrix tends to be ill-conditioned due to noise in sampling.
At a high sample count, the direct matrix approach outperforms SGD due to its ability to find a global minimum. 
For a high-order polynomial (order 7),  the direct matrix approach becomes computationally expensive whereas SGD runs at at least $2\times$ faster speed while giving similar relMSE.

\section{Discussion}

\begin{figure}[t!]
    \centering

\definecolor{contour}{RGB}{60,60,60}

\newcommand{\PlotSplitVerticalThree}[3]{
    \begin{scope}
        \clip (0,0) -- (0.0,4.5) -- (2.83,4.5) -- (2.83,0) -- cycle;
        \path[fill overzoom image=figures/#1] (0,0) rectangle (8.5cm,4.5cm);
    \end{scope}
    \begin{scope}
        \clip (2.83,0) -- (2.83,4.5) -- (5.66,4.5) -- (5.66,0) -- cycle;
        \path[fill overzoom image=figures/#2] (0,0) rectangle (8.5cm,4.5cm);
    \end{scope}
    \begin{scope}
        \clip (5.66,0) -- (5.66,4.5) -- (8.5,4.5) -- (8.5,0) -- cycle;
        \path[fill overzoom image=figures/#3] (0,0) rectangle (8.5cm,4.5cm);
    \end{scope}
    \draw[draw=white,thick] (2.83,0.0) -- (2.83,4.5);
    \draw[draw=white,thick] (5.66,0) -- (5.66,4.5);
}


\small
\setlength\tabcolsep{1pt}
\renewcommand{\arraystretch}{0.5} 
\begin{tabularx}{\linewidth}{c@{}}
    \adjustbox{valign=c}{\begin{tikzpicture}[scale=0.95]
        \PlotSplitVerticalThree{incremental/inc_bathroom_mce_256_mrse}{incremental/inc_bathroom_sgd_online_o3_256_mrse}{incremental/inc_bathroom_sgd_o3_256_mrse}
        \begin{scope}
            \filldraw[white,ultra thick] (2.83/2, 4.5) circle (0pt) node[anchor=north,rotate=0]
            	{\fontsize{7.8}{5}\selectfont\Outlined[white][contour]{MC}};
            \filldraw[white,ultra thick] (2.83/2, 0.0) circle (0pt) node[anchor=south,rotate=0]
            	{\fontsize{7.8}{5}\selectfont\Outlined[white][contour]{relMSE: $1.83$}};
			\filldraw[white,ultra thick] (2.83+2.83/2, 4.5) circle (0pt) node[anchor=north,rotate=0]
            	{\fontsize{7.8}{5}\selectfont\Outlined[white][contour]{GD - O3 - incremental}};            
            \filldraw[white,ultra thick] (2.83+2.83/2, 0.0) circle (0pt) node[anchor=south,rotate=0]
            	{\fontsize{7.8}{5}\selectfont\Outlined[white][contour]{relMSE: $1.63$}};
            \filldraw[white,ultra thick] (2.83*2+2.83/2, 4.5) circle (0pt) node[anchor=north,rotate=0]
            	{\fontsize{7.8}{5}\selectfont\Outlined[white][contour]{GD - O3 - T=1}};            
            \filldraw[white,ultra thick] (2.83*2+2.83/2, 0.0) circle (0pt) node[anchor=south,rotate=0]
            	{\fontsize{7.8}{5}\selectfont\Outlined[white][contour]{relMSE: $1.56$}};
            	
			\begin{scope}[shift={(8.5-0.5,0.0)}]            
            	\path[fill stretch image=figures/map] (0,0) rectangle (0.5,2.0); 
        	\filldraw[white,ultra thick] (0.25, 0.0) circle (0pt) node[anchor=south,rotate=0]
            	{\fontsize{6}{5}\selectfont\Outlined[white][contour]{0.0}};
        	\filldraw[white,ultra thick] (0.25, 2.0) circle (0pt) node[anchor=north,rotate=0]
            	{\fontsize{6}{5}\selectfont\Outlined[white][contour]{0.05}};
        	\filldraw[white,ultra thick] (0.025, 1.5) circle (0pt) node[anchor=north east,rotate=90]
            	{\fontsize{6}{5}\selectfont\Outlined[white][contour]{relMSE}};
            \end{scope}
        \end{scope}
    \end{tikzpicture}}
\end{tabularx}

\undefinecolor{contour}
    \caption{
        Incremental regression-based Monte Carlo integration with gradient descent. We render the \textsc{Bathroom} scene (256 spp) and compare with \ac{MC} and non-incremental gradient descent estimator. While the incremental gradient descent has slightly higher error than its non-incremental counterpart, both achieves less error than \ac{MC} integration. relMSE values are scaled by $10^{-2}$.
    }
    \label{fig:inc_exp}
\end{figure}
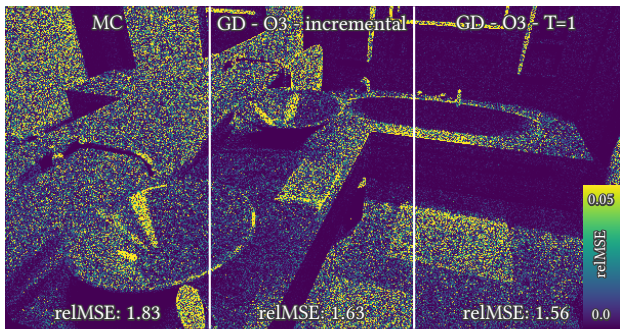

\paragraph{Incremental estimators}
In our implementation, we assume that the $N$ sample estimator $\CVEstimator$ is constructed by first finding the solution to regression $\CVfunctionPSS$ using all the $N$ samples, and then evaluating $\CVEstimator$ with the same $N$ samples. 
Therefore, when $N$ is changed to $N+1$, we need to solve regression again with $N+1$ samples first and then re-evaluate $\CVEstimator$ with the new $\CVfunctionPSS$ using $N+1$ samples.
In some applications, such as progressive rendering, being able to incrementally update both $\CVEstimator$ and $\CVfunctionPSS$ as we add a new sample is desirable.
Thanks to the use of control variates, our formulation allows such incremental update as follows.

Let us denote $\CVEstimator_N$ and $\CVfunctionPSS_N$ as the estimated value and the solution to regression with $N$ samples. 
Using an online regression algorithm, one can obtain $\CVfunctionPSS_{N+1}$ based on $\CVfunctionPSS_N$ by adding a new sample $u_{N+1}$. 
We define the incremental estimate $\CVEstimator_{N+1}$ as 
\begin{align}
\resizebox{0.9\columnwidth}{!}{
$\CVEstimator_{N+1} = \frac{1}{N+1} \left(N \CVEstimator_{N} + \left( G_{N} + \IntegrandPSS(u_{N+1}) - \CVfunctionPSS_{N}(u_{N+1}) \right)\right)$
}
\end{align}     
where $G_{N+1}$ is the integral of $\CVfunctionPSS_{N+1}$. 
This estimator remains unbiased since  
\begin{align}
   & \ExpectedOp{\CVEstimator_{N+1}} \nonumber\\
  & = \ExpectedOp{\frac{1}{N+1} \left(N \CVEstimator_N + \left( G_{N} + \IntegrandPSS(u_{N+1}) - \CVfunctionPSS_{N}(u_{N+1}) \right)\right)} \nonumber\\
    &=\frac{1}{N+1} \left(N \Integral + \ExpectedOp{\left( G_{N} + \IntegrandPSS(u_{N+1}) - \CVfunctionPSS_{N}(u_{N+1}) \right)}\right) \nonumber\\
   & = \frac{1}{N+1} \left(N \Integral +\Integral \right) = \Integral.
\end{align}
The expected squared error of incremental estimators is likely worse than that of non-incremental counterparts; the $N$ sample incremental estimator now combines the results of regression for all the $N-1, N-2, ..., 1$ samples which might be numerically unstable at the beginning.
The incremental estimators are, however, often computationally less expensive for the same number of samples than the non-incremental counterparts.
We leave the exact analysis of these differences as future work, but we show a preliminary result in Fig.~\ref{fig:inc_exp} where the incremental estimate is performed with the gradient descent estimator. The numerical error of the incremental estimate is slightly higher than the non-incremental version, but it still outperforms \ac{MC} integration.

\paragraph{Differences from MCLS}

\sloppy
Our regression-based Monte Carlo integration bears some similarities to Monte Carlo Least-Squares MCLS)~\citep{nakatsukasa2018approximate}.

\citeauthor{nakatsukasa2018approximate} reformulated Monte Carlo integration as least-squares regression followed by the analytical integration of the resulting function. 
MCLS also performs no worse than Monte Carlo integration as long as a constant function is included in regression. 
There are two major differences between MCLS and our approach.
Firstly, MCLS considers a model function $g$ which integrates exactly to $F$ (i.e., $F = G$).

Such a family of model functions is far more restrictive than our model function which is allowed to integrate to \emph{any} value.
Secondly, \citeauthor{nakatsukasa2018approximate} claims that MCLS is derived under an entirely different framework than control variates, thus its connection to control variates is unclear.
In contrast, we show how control variates is closely related to the combination of Monte Carlo integration and least-squares regression. 

When we use a model function $\CVfunctionPSS$ such that $\CVIntegral = \Integral$ (i.e., keeps the integral value), then our approach in fact reduces to MCLS. 
MCLS is thus can be seen as a special case of our formulation.

\paragraph{Russian roulette}
In our current implementation, we manually set all path lengths to be the same. 
This is because, for regression it is necessary to fix the dimensionality of the search space.
Russian roulette, however may result in paths of different length/dimensionality.

One simple way to combine Russian roulette with our approach would be to perform regression up to a certain prescribed dimension and only invoke Russian roulette for paths that have dimensionality higher than the prescribed one. 
The high-dimensional sample components can then be projected back into the lower-dimensional space.

\section{Limitations and future work}

While our method often just works as a drop-in improvement over Monte Carlo integration in many cases, there are several limitations and future work.

Regression in our method incurs additional computation over MC integration. 
Regression may or may not be worth the additional cost depending on the amount of error reduction in the end. 
In general, regression becomes computationally more costly for a more complex model function, but a complex function typically leads to better error reduction. 
There is a trade-off among the reduction of error, the complexity of a model function, and the computational overhead of the regression process. 
Finding an optimal balance among those can potentially be achieved in an adaptive manner.
Our formulation does not specify how regression should be solved in practice, so there is a large room of design choices in its implementation.

Since our formulation is based on the primary sample space, applying our method to a higher number of bounces in light transport simulation is challenging.
The dimensionality of the primary sample space is directly proportional to the number of bounces, and regression in a high dimensional is not numerically stable. 
It is being recognized that light transport simulation has a much lower dimensional structure in the original path space, as was utilized by path guiding methods~\cite{Vorba:2014:Online,Muller:2017:Practical}.
It would be interesting to study how our regression-based formulation would benefit from such a low dimensional structure of light transport.

Depending on how the regression is implemented, our method can have a very small bias. 
When the same set of samples are used for both regression of the control variate and estimation of the difference, our estimator can be biased.
\citet{Owen:2000:Safe} and \citet{hickernell2005control} have analyzed 
the amount of bias introduced when estimating the $\alpha$ parameter and evaluating the difference estimator using the same set of samples in control variates. 
In brief, the bias is negligible compared to the variance of the estimator and reduces at the order of $O(N^{-1})$ for $N$ samples as opposed to $O(N^{-1/2})$ of the error of Monte Carlo integration.
This makes our estimator \emph{consistent} and the bias reduces faster than variance.
One can also always remove this bias by separating a set of samples into two halves, performing regression only on the other half, and taking the average of the two difference estimators, in a manner akin to cross-validation. 
We, however, have not used this approach in the results in this paper since we have observed that bias is negligible.
Our analysis is not influenced by this bias since this bias happens only when the same set of samples are used in its implementation of regression, while our analysis does not assume any specific implementation of regression.

\begin{acks}
We thank the anonymous reviewers for their comments that helped shape the paper. We thank the following for scenes used in our experiments: Wing42 (\textsc{Dinning-room}), nacimus (\textsc{Bathroom}), Karl Li (\textsc{PBRT-book}), julioras3d (\textsc{chopper-titan}), Greyscalegorilla (\textsc{vw-van}), Benedikt Bitterli (\textsc{Teapot}), MtChimp2313 (\textsc{House}).
This project is supported in part by the Natural Sciences and Engineering Research Council of Canada (NSERC) RGPIN1507.
\end{acks}

\bibliographystyle{ACM-Reference-Format}
\bibliography{main}

\end{document}